\documentclass[11pt]{article}
\usepackage{booktabs}       % professional-quality tables
\usepackage{amsfonts}       % blackboard math symbols
\usepackage{amsmath}
\usepackage{graphicx}
\usepackage{amsthm}
\usepackage{amssymb}
\usepackage{multirow}
\usepackage{makecell}
\usepackage{algorithmic}
\usepackage{algorithm}
\usepackage[round]{natbib}
\usepackage[colorlinks,linkcolor=magenta,filecolor=blue,citecolor=blue,urlcolor=blue]{hyperref}%
\usepackage[top=1in, left=1in, right=1in, bottom=1in]{geometry}
\usepackage{subfigure}
\usepackage{amssymb}% http://ctan.org/pkg/amssymb
\usepackage{pifont}
\usepackage{mathtools}
\usepackage{stmaryrd}
\usepackage[T1]{fontenc}

\newtheorem{theorem}{Theorem}
\newtheorem{corollary}{Corollary}
\newcommand{\floor}[1]{\lfloor #1 \rfloor}

\allowdisplaybreaks

\newtheorem*{Lemma*}{Lemma}
\newtheorem*{Theorem*}{Theorem}
\newtheorem*{Corollary*}{Corollary}

\newcommand{\beq}{\begin{equation}}
\newcommand{\eeq}{\end{equation}}

\begin{document}

\title{\bf Communication-Efficient TeraByte-Scale Model Training Framework for Online Advertising}

\author{Weijie Zhao$^1$, Xuewu Jiao$^2$, Mingqing Hu$^2$, Xiaoyun Li$^1$, Xiangyu Zhang$^2$, Ping Li$^1$\\
$^1$ Cognitive Computing Lab, Baidu Research\\
$^2$ Baidu Search Ads (Phoenix Nest), Baidu Inc. \\
10900 NE 8th St. Bellevue, Washington 98004, USA\\
No. 10 Xibeiwang East Road, Beijing 100193, China\\
\small \{weijiezhao jiaoxuewu humingqing, xiaoyunli, zhangxiangyu06, liping11\}@baidu.com
}
%\affiliation{%
%  \institution{Baidu Inc.}
%\email{{weijiezhao,jiaoxuewu,humingqing,v_lixiaoyun02,zhangxiangyu06,liping11}@baidu.com}

\date{\vspace{0.5in}}

\maketitle

\begin{abstract}\vspace{0.1in}
\noindent Click-Through Rate (CTR) prediction is a crucial component in the online advertising industry. In order to produce a personalized CTR prediction, an industry-level CTR prediction model commonly takes a high-dimensional (e.g., 100 or 1000 billions of features) sparse vector (that is encoded from query keywords, user portraits, etc.) as input. As a result, the model requires Terabyte scale parameters to embed the high-dimensional input. Hierarchical distributed GPU parameter server has been proposed to enable GPU with limited memory to train the massive network by leveraging CPU main memory and SSDs as secondary storage.
We identify two major challenges in the existing GPU training framework for massive-scale ad models and propose a collection of optimizations to tackle these challenges: (a) the GPU, CPU, SSD rapidly communicate with each other during the training. The connections between GPUs and CPUs are non-uniform due to the hardware topology. The data communication route should be optimized according to the hardware topology; (b) GPUs in different computing nodes frequently communicates to synchronize parameters. We are required to optimize the communications so that the distributed system can become scalable.\\

\noindent In this paper, we propose a hardware-aware training workflow that couples the hardware topology into the algorithm design.
To reduce the extensive communication between computing nodes, we introduce a $k$-step model merging algorithm for the popular Adam optimizer and provide its convergence rate in non-convex optimization. To the best of our knowledge, this is the first application of $k$-step adaptive optimization method in industrial-level CTR model training. The numerical results on real-world data confirm that the optimized system design considerably reduces the training time of the massive model, with essentially no loss in accuracy.\\

\noindent All experiments are conducted on Baidu's ads production system which is part of the PaddlePaddle deep learning platform. \url{https://www.paddlepaddle.org.cn}
\end{abstract}

\newpage

\section{Introduction}

Online advertising industry displays sponsored ads in the searching results and recommendations, and profits from the ads publisher whenever the displayed ad is clicked. Baidu (\url{www.baidu.com}) has developed over the past two decades a mature and powerful advertising system (a.k.a. ``Phoenix Nest'') that makes a handsome profit for the corporation. As reviewed in (e.g.,)~\cite{Proc:Fan_KDD19, Proc:Zhao_MLsys20}, Baidu started to use distributed parameter servers for massive scale logistic regression around 2010 and launched their distributed deep learning training system in production around 2013. An important task in the ads system is to train machine learning models for Click-Through-Rate (CTR) prediction~\citep{Article:Broder_2002,
fain2006sponsored}. CTR measures the number of click-throughs per ad display---it is employed to quantify the performance of an ad display strategy. Thus, CTR prediction becomes one of the most crucial components in the online advertising industry, in which the prediction accuracy directly affects the revenue. In recent years, at Baidu Search Ads, multiple directions has been taken to improve the performance of CTR prediction, including using approximate near neighbor  search for improving recalls~\citep{Proc:Fan_KDD19,Proc:Zhou_NeurIPS19}, GPU-based ads production systems~\citep{Proc:Zhao_CIKM19,Proc:Zhao_MLsys20}, model compression~\citep{Proc:Xu_SIGMOD21}, etc.

% Online advertising industry displays sponsored ads in the searching results and recommendations, and profits from the ads publisher whenever the displayed ad is clicked. Therefore, the ads system targets to allocate the ads that are likely to be clicked---the ads should be relevant to the user's search keywords, his/her personal interests, etc. Click-Through-Rate (CTR) measures the number of click-throughs per ad display---it is employed to quantify the performance of an ad display strategy. CTR prediction~\cite{Article:Broder_2002,
% fain2006sponsored,Proc:Fan_KDD19} becomes one of the most crucial components in the online advertising industry, whose accuracy directly affects the revenue.

In the early stage of this line of research, data scientists used classical regression algorithms, e.g., logistic regression and gradient boosting, for CTR prediction~\citep{edelman2007internet,graepel2010web}.
In recent years, deep learning, which could achieve state-of-the-art performance in numerous tasks, has attracted great attention in both industry and academia. With the rapid development of computing resources (e.g. GPU), machine learning researchers have applied large models with a massive number of parameters in practice to achieve higher learning accuracy~\citep{shan2016deep,covington2016deep,zhou2018deep,qu2016product}. For CTR prediction, in this paper, we consider  extremely sparse input which encodes the query keywords, user portrait, etc. The sparse input vector is first embedded into a low-dimensional ($\sim100$) dense representation and then fed to subsequent network components, i.e., attention component and multi-layer perceptron. Because of the ultra-high dimensionality of the input layer, the CTR model requires a massive number of parameters~($>$10~TB).

Besides achieving high accuracy, it is also important for practitioners to be able to train the large CTR model in an efficient manner. This is because: 1) the CTR prediction model should be capable of incorporating the continuously changing information; 2) machine learning researchers may need to rapidly train CTR models for evaluation and strategy development. In order to tackle the massive-scale learning problem, in industrial applications, people commonly resort to a distributed parameter server~\citep{valiant1990BSP,ho2013more,FlexPS} across a computing cluster, with dozens to hundreds of nodes. However, training in a large computing cluster requires significant communications and synchronizations that limit the efficiency of the training task. Recently, to reduce the communications, distributed GPU parameter servers~\citep{Proc:Zhao_MLsys20} are introduced to train massive-scale models on a handful of nodes accelerated by GPUs and SSDs.

Nonetheless, previous studies focus on effectively using CPU main memory and SSDs to handle out-of-GPU-memory scenarios. When the model becomes more complicated and requires more computing nodes, the communication issue will still exist and slow down the training process.

\begin{figure}[htbp]
\centering
\includegraphics[width=.5\textwidth]{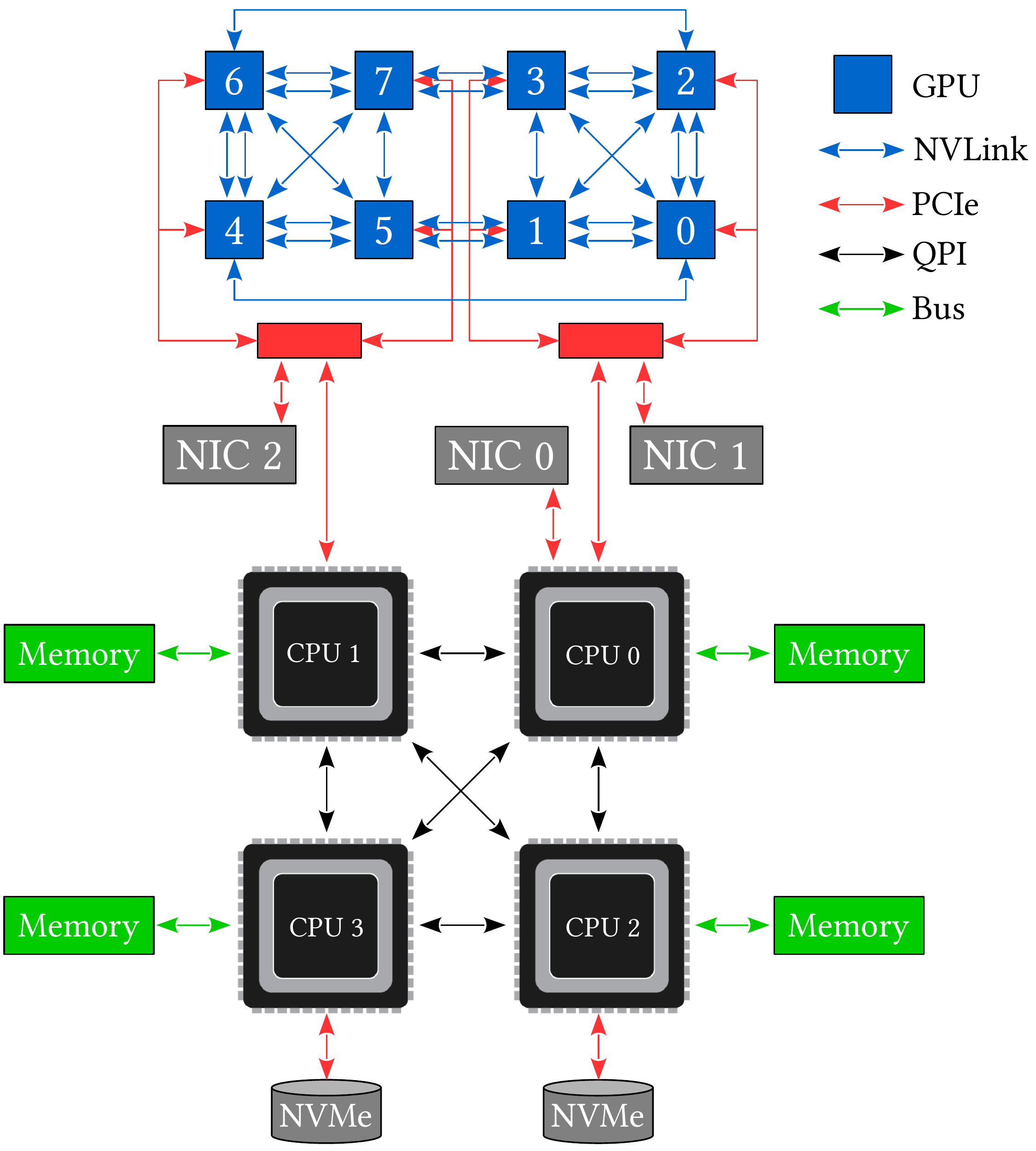}
\caption{An example hardware topology of a GPU node.}
%\vspace{-.5cm}
\label{fig:topology}
\end{figure}

\vspace{0.05in}
\textbf{Challenges \& approaches.}
We identify two major challenges in the design of GPU-SSD accelerated training framework: local communication inside a computing node and network communications among nodes.
First, GPUs in the same computing node are required to rapidly communicate with each other to sum up the gradients and synchronize the parameters. Thus, the speed of transmitting gradients/parameters become crucial. However, it is non-trivial to perform the collective reduction/broadcast in a real system because of the hardware topology: the bandwidths between GPUs are not identical. For example, in Figure~\ref{fig:topology}, GPU 0, 1, 2, and 3 are on one socket while the other 4 GPUs (4, 5, 6, and 7) locate on the other socket.
Since GPU 1, 3, 5, and 7 need to connect the GPUs across two sockets, their bandwidths to the GPUs in their own socket are reduced: e.g., the link between GPU 0 and 2 has 2X bandwidth than the link between GPU 1 and 3. Even in the same socket, the bandwidths between two GPUs are now the same: e.g., GPU 0 and GPU 3 have only one NVLink bridge connector between them so that it has only half bandwidth  than the one between GPU 0 and 2. A standard collective reduction/broadcast is not aware of the hardware topology and thus yields suboptimal performance. We allocate the I/O workload onto the CPU cores that are co-located to the I/O device to reduce the data transferring across NUMA sockets. Also, GPU, CPU, and SSDs in the same node rapidly communicate during the training to load/cache/dump sparse parameters. The communications are required to be optimized so that the inner-node I/O does not become a bottleneck of the training framework.

The second challenge is the communication between the computing nodes. Although GPU RDMA is discussed in~\cite{Proc:Zhao_MLsys20}, the inter-node communication through the network is slow and has incomparable bandwidth with the inner-node PCI-E data transfer---frequent inter-node synchronizations hurt the training performance. We introduce a $k$-step model merging algorithm with Adam~\citep{Proc:Kingma_ICLR15} as the local optimizer, which is different from the common local SGD ($k$-step SGD) in that Adam employs adaptive coordinate-wise learning rates during the training process.

\newpage

%\vspace{0.05in}
\textbf{Contributions.} Specifically, our contributions include:
\begin{itemize}
\item We propose a hardware-aware training workflow that couples the hardware topology into the algorithm design.
The proposed techniques have been implemented in our real-world ads system and obtained a good success in our company. We believe these engineering details would be interesting to the machine learning community.

\item To reduce the extensive communication between computing nodes, we introduce a $k$-step model merging algorithm for Adam and provide its convergence rate in non-convex optimization. To the best of our knowledge, this is the first application of $k$-step adaptive optimization in industrial-level CTR model training.

%\item We present a page cache management policy to enable direct disk I/O. The SSD I/O time is significantly reduced with this optimization.

\item We experimentally evaluate the proposed system on real-world ads data. The empirical results confirm the effectiveness of our proposed optimizations.
\end{itemize}

\section{Preliminaries}

In this section, we introduce the CTR prediction model and the distributed GPU parameter server. Both concepts are the foundations of the proposed framework in this paper.

\subsection{CTR Prediction Model}

\begin{figure}[htbp]
\centering
\includegraphics[width=.57\textwidth]{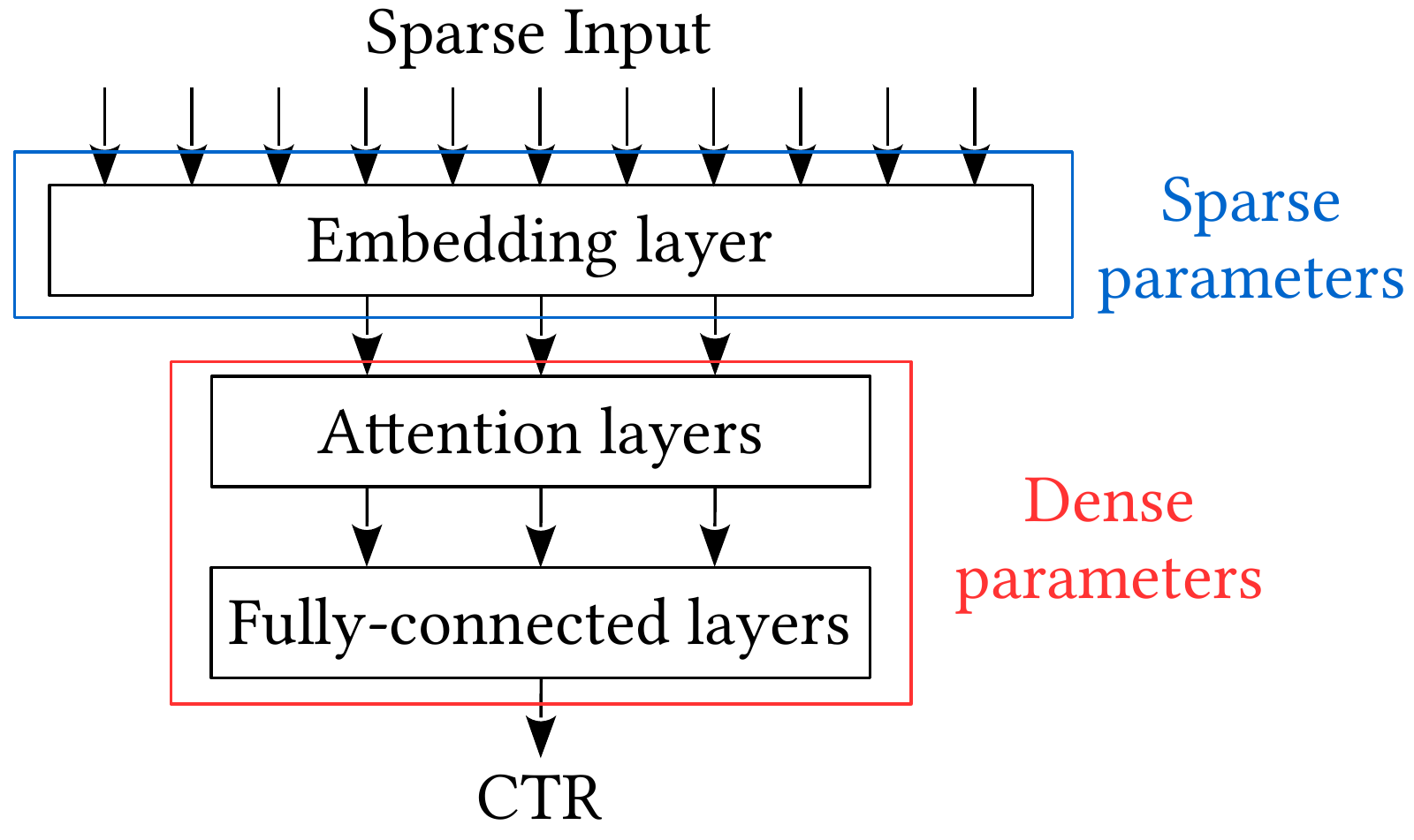}
\caption{An example structure of our CTR prediction model.}
\label{fig:ctr}
\end{figure}

Given an ad, we target to use a CTR prediction model to predict the probability that the ad will be clicked based on a variety of input sources, e.g., search keywords, user portrait, and more ad features.
Those input sources are commonly encoded in a one-hot/multi-hot representation---the input vector is formalized as an extremely sparse vector with $\sim$$10^{11}$ dimensions and only $\sim$$100$ non-zeros on average.
Figure~\ref{fig:ctr} depicts an example neural network architecture of a commercial CTR prediction model.
The ultra-high dimensional vector is fed to an embedding layer to produce a low-dimensional representation. After that, more complicated neural network components, such as attention and fully-connected layers~\citep{shan2016deep,cheng2016wide,covington2016deep,qu2016product,guo2017deepfm,lian2018xdeepfm}, are employed on the low-dimensional embedding to predict the CTR.

\subsection{Why Not Using Hashing?}

This is the same question the authors were asked over and over in the past years. We have answered it thoroughly in~\cite{Proc:Zhao_MLsys20}, that commercial search engines are unable to tolerate any noticeable (e.g., $0.1\%$) loss of accuracy which directly impact the revenues.

Nevertheless, here we present the same Table~\ref{tab_FengCao} taken from~\cite{Proc:Zhao_MLsys20}, for reporting the previous efforts (since 2015) in developing hashing methods for online ads system for web search. As we can see, the baseline deed learning model (``Baseline DNN'') achieved an AUC of 0.7670. Although hashing substantially reduced the model size, it is not realistic to replace the current production system because the loss of accuracy is not acceptable, even though the loss appears very small from academic perspective.

\begin{table}[ht]
\caption{OP+OSRP hashing for Web Search Sponsored Ads Data, taken from~\cite{Proc:Zhao_MLsys20} }
\begin{center}{
\small
{\begin{tabular}{l r c c c c c }
\hline \hline
       &\# Nonzero Weights  & Test AUC\\
\hline
Baseline LR &199,359,034,971 &0.7458 \\
Baseline DNN &  &0.7670\\
Hash+DNN (k = $2^{32}$) & 3,005,012,154 &0.7556\\
Hash+DNN (k = $2^{31}$) & 1,599,247,184 &0.7547\\
Hash+DNN (k = $2^{30}$) & 838,120,432 &0.7538\\
Hash+DNN (k = $2^{29}$) & 433,267,303 &0.7528\\
Hash+DNN (k = $2^{28}$) & 222,780,993 &0.7515\\
Hash+DNN (k = $2^{27}$) & 114,222,607 &0.7501\\
Hash+DNN (k = $2^{26}$) & 58,517,936 &0.7487\\
Hash+DNN (k = $2^{24}$) & 15,410,799 &0.7453\\
Hash+DNN (k = $2^{22}$) & 4,125,016&0.7408
\\\hline\hline
\end{tabular}}
}
\end{center}\label{tab_FengCao}
\vspace{-0.1in}
\end{table}

\subsection{Distributed Hierarchical GPU Parameter Server}
Currently, the off-the-shelf GPU with the largest memory capacity has 40 GB High-Bandwidth Memory (HBM). Comparing with the 10 TB parameters of CTR models, GPUs have too limited HBM to hold the massive-scale parameters. To tackle this challenge, distributed hierarchical GPU parameter server is introduced in~\cite{Proc:Zhao_MLsys20}. The key observation made this possible is that: the input is extremely sparse (only hundreds of non-zeros)---only hundreds of parameters in the massive-scale embedding layer are referenced and required to be loaded during the feed forward and back propagation. Therefore, the working parameters in a mini-batch fit in the GPU memory. \cite{Proc:Zhao_MLsys20} maintains a distributed hash table in the HBM across GPUs to store the working parameters and use CPU main memory and NVMe SSDs as secondary storage.

%%%%%%%%%%%%%%%%%%%%%%%%%%%%%%%%%%%%%%%%%%%%%%%
\begin{algorithm}[htbp]
\caption{Training Workflow}
\label{alg:workflow}
\algsetup{linenodelimiter=.}
\begin{algorithmic}[1]
\WHILE{\textit{not converged}}
	\STATE \textit{batch} $\leftarrow$ \textit{read\_batch\_from\_network}() \label{line:hdfs}
	\STATE \textit{working}$\leftarrow$\textit{pull\_parameters}(\textit{batch})~\label{line:pull-parameters}
	\STATE {\small \textit{minibatches} $\leftarrow$ \textit{shard\_batch}(\textit{batch,\#GPU,\#minibatch})}\hspace*{-3cm}\label{line:part-transfer-begin}
	\STATE \textit{part} $\leftarrow$ {\small{\textit{partition\_parameters\_to\_GPUs}}(\textit{working},\#GPU)}\hspace*{-3cm}
	\FOR{$i \leftarrow1$ \textbf{to} \textit{\#GPU}}
		\STATE \textit{transfer\_to\_GPU}($i$,$\textit{minibatches}_{i}$)
		\STATE \textit{insert\_into\_hash\_table}($i$,$\textit{part}_{i}$)
	\ENDFOR \label{line:part-transfer-end}
	\FOR{$j \leftarrow 1$ \textbf{to} \textit{\#minibatch}}
		\STATE \textit{pull\_peer\_GPU\_kernel}($\textit{minibatch}_{j}$) \label{line:gpu-pull}
		\STATE 	$\Delta_{j}$ $\leftarrow$ \textit{train\_mini-batch\_kernel}($j$) \label{line:train}
		\STATE 	\textit{push\_parameters\_updates\_back\_kernel}($\Delta_{j}$) \label{line:gpu-push}
	\ENDFOR
	\STATE $\Delta$ $\leftarrow$ \textit{pull\_updates\_peer\_GPU}() \label{line:hbm-collect}
	%\STATE \textit{push\_remote\_MEM-PS}($\Delta$) \label{line:push-remote}
	\STATE \textit{parameters\_to\_dump} $\leftarrow$ \textit{update\_local\_cache}() \label{line:dump-begin}
	\STATE \textit{push\_local\_SSD}(\textit{parameters\_to\_dump}) \label{line:dump-end}
\ENDWHILE
\end{algorithmic}
\end{algorithm}%\vspace{-0.in}
%%%%%%%%%%%%%%%%%%%%%%%%%%%%%%%%%%%%%%%%%%%%%%%

\vspace{0.1in}
\noindent\textbf{Training workflow.}
Algorithm~\ref{alg:workflow} illustrates the distributed hierarchical parameter server training workflow. The training data batches are streamed into the main memory through the network card (NIC 0 in Figure~\ref{fig:topology}) with a network file system, e.g., HDFS (line~\ref{line:hdfs}). The parallel training framework falls in the data-parallel category~\citep{li2014scaling,cui2014exploiting,Geeps,luo2018parameter}. Each computing node processes its own training batches---different nodes receive different training data from HDFS. The streamed data for different nodes are in an i.i.d. distribution.
Then, each node identifies the union of the referenced parameters in the current received batch and pulls these parameters from the local memory/SSD and its peer computing nodes (line~\ref{line:pull-parameters}). The local in-memory cache management loads the local parameters stored on SSDs into the memory and requests other nodes for the remote parameters through the network.
After we load all the referenced parameters into the memory, these parameters are then partitioned and transferred to the HBM in GPUs. The parameters are partitioned in a non-overlapped fashion to effectively utilize the limited GPU memory. When a GPU worker thread requires the parameter on another GPU, it fetches the parameter from the remote GPU and pushes the updates back to the remote GPU through high-speed inter-GPU hardware connection NVLink~\citep{foley2017ultra}.
Then, the data batch is partitioned into multiple mini-batches and fed to each GPU worker thread (line~\ref{line:part-transfer-begin}-\ref{line:part-transfer-end}).
An inter-GPU parameter synchronization is performed after each mini-batch.
Each training worker pulls the required parameters of its corresponding mini-batch from its peer GPUs (line~\ref{line:gpu-pull}), performs feed forward and backward propagation to update the parameters (line~\ref{line:train}), and interacts with its peer GPUs to update the referenced parameters on other GPUs (line~\ref{line:gpu-push}).
A collection of CPU threads collects the updates from the GPU (line~\ref{line:hbm-collect})
%, pushes the remote parameter updates to the remote peer HBM-PS (line~\ref{line:push-remote}),
and dumps infrequently used parameters to the SSDs when the main memory usage reaches its capacity (line~\ref{line:dump-begin}-\ref{line:dump-end}). GPU HBM, CPU main memory, and SSDs communicate in a hierarchical storage fashion. The upper-level module acts as a high-speed cache of the lower-level module. %Moreover, the 3-level parameter server is specially optimized for front/back propagation in deep learning scenarios.

\section{Hardware-Aware Training Framework}

\subsection{Core Binding}
As shown in Figure~\ref{fig:topology}, the physical topology of the system determines the data access path between two devices: GPU 0-3 are connected with the same PCIe slot/switch to CPU 0 while GPU 4-7 are connected to CPU 1; CPU 3 and 4 have connected to an NVMe SSD, respectively. As a result, CPU 0 and 1 have faster access to the GPUs and CPU 2 and 3 have faster SSD access.
The 4 CPUs are installed in 4 sockets as the NUMA fashion. The communications between CPUs are transferred through QPI.
CPU 2 and 3 connect two 4 NVMe SSDs, respectively. We allocate one physical CPU core for each SSD to manage the I/O: these cores are responsible for loading and dumping parameters from/to SSDs. The loaded parameters are scattered into the memory of all 4 CPUs---the NUMA memory policy is set to \texttt{MPOL\_INTERLEAVE} that interleaves page allocations across the CPUs. A network card (NIC 0) is attached to CPU 0 via PCIe to read training data from network file systems. A core in CPU 0 is assigned to solely perform the network I/O. NIC 1 and 2 are controlled by GPUs on the corresponding PCIe switch to perform RDMA GPUDirect I/O with GPUs on other computing nodes. The RDMA communication bypasses the involvement of CPU and main memory access so that we can achieve a larger bandwidth and lower latency.

\subsection{Two-Phase GPU Communications}
During the training workflow described in Algorithm~\ref{alg:workflow}, GPUs in the same node rapidly communicates with each other to pull referenced parameters and push update parameters in the distributed hash table across GPUs---the hash table stores the sparse parameters of the embedding layer. However, due to the hardware topology (Figure~\ref{fig:topology}), not all GPUs have direct connections: e.g., GPU 0 cannot directly communicate with GPU 5. In conventional practice, this cross socket communications takes the following route: GPU 0, PCIe switch, CPU0 (memory), QPI, CPU 1 (memory), PCIe switch, and GPU 5. The communication route is inefficient since it involves multiple data copies between CPUs and GPUs.

\begin{figure}[htbp]
\centering
\includegraphics[width=.4\textwidth]{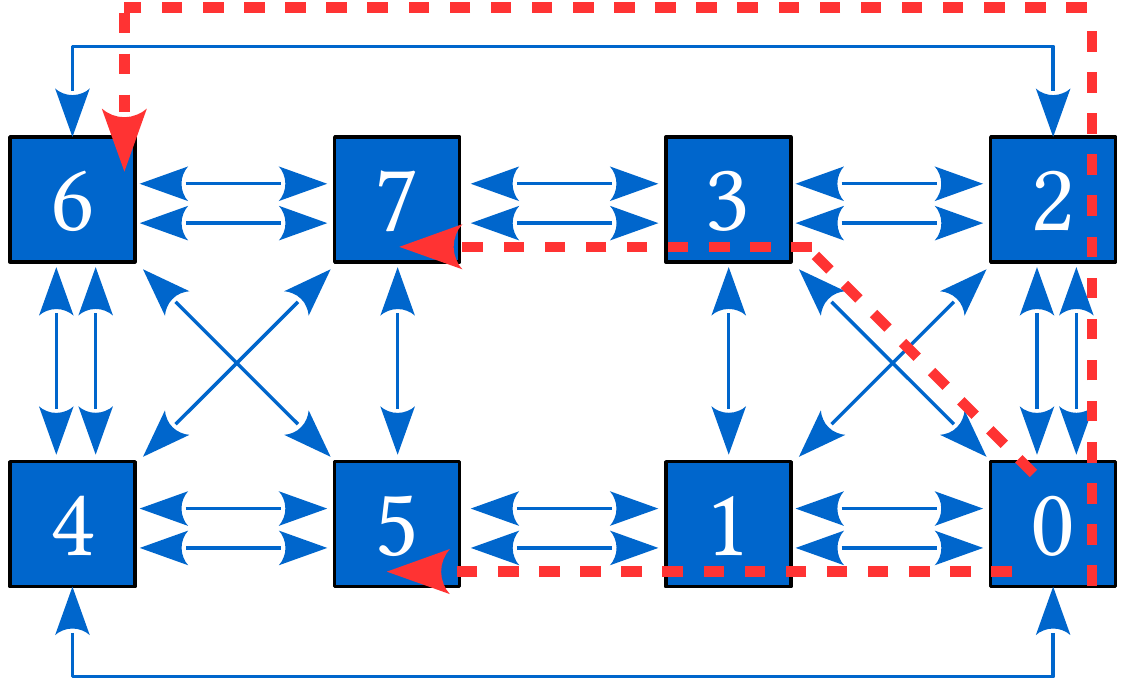}
\caption{Example two-phase GPU communication patterns.}
\label{fig:2step}
\end{figure}

We propose a two-phase communication strategy to tackle this challenge. The CPU-involved GPU communications only exist when we try to perform cross PCIe socket/switch GPU data transfer without NVLink connection. Figure~\ref{fig:2step} depicts three representative communication patterns from GPU 0. The communication patterns from other GPUs are similar and thus are omitted. GPU 0 has no NVLink connection with GPU 5, 6, and 7. Instead of routing through CPUs with PCIe switch, we allocate a buffer on GPU 1, 2, and 3, respectively, to forward the data from GPU 0 to 5, 6, and 7. The data transfer route using a middleman GPU to forward the communication eliminates the involvement and data copy of CPU and PCIe switch---the two-phase GPU communications only transfer data through NVLink which has a larger bandwidth.

%\subsection{Fused Communications}

\subsection{SSD Direct I/O}
By default, common operating systems enable dirty page write back that periodically flush the modified cached pages into the disk. Moreover, a disk cache is also maintained by the operating system in the memory to accelerate the disk access---it always probes the in-memory page cache before accessing disks. However, the hierarchical distributed GPU parameter server has already kept a page cache of parameters. In this case, the system disk cache introduces an unnecessary memory probing overhead. Thus, we disable the dirty page write back and use non-buffered direct disk I/O to access SSDs.

\section{$k$-Step Model Merging}

\begin{figure}[htbp]
\centering
\includegraphics[width=.5\textwidth]{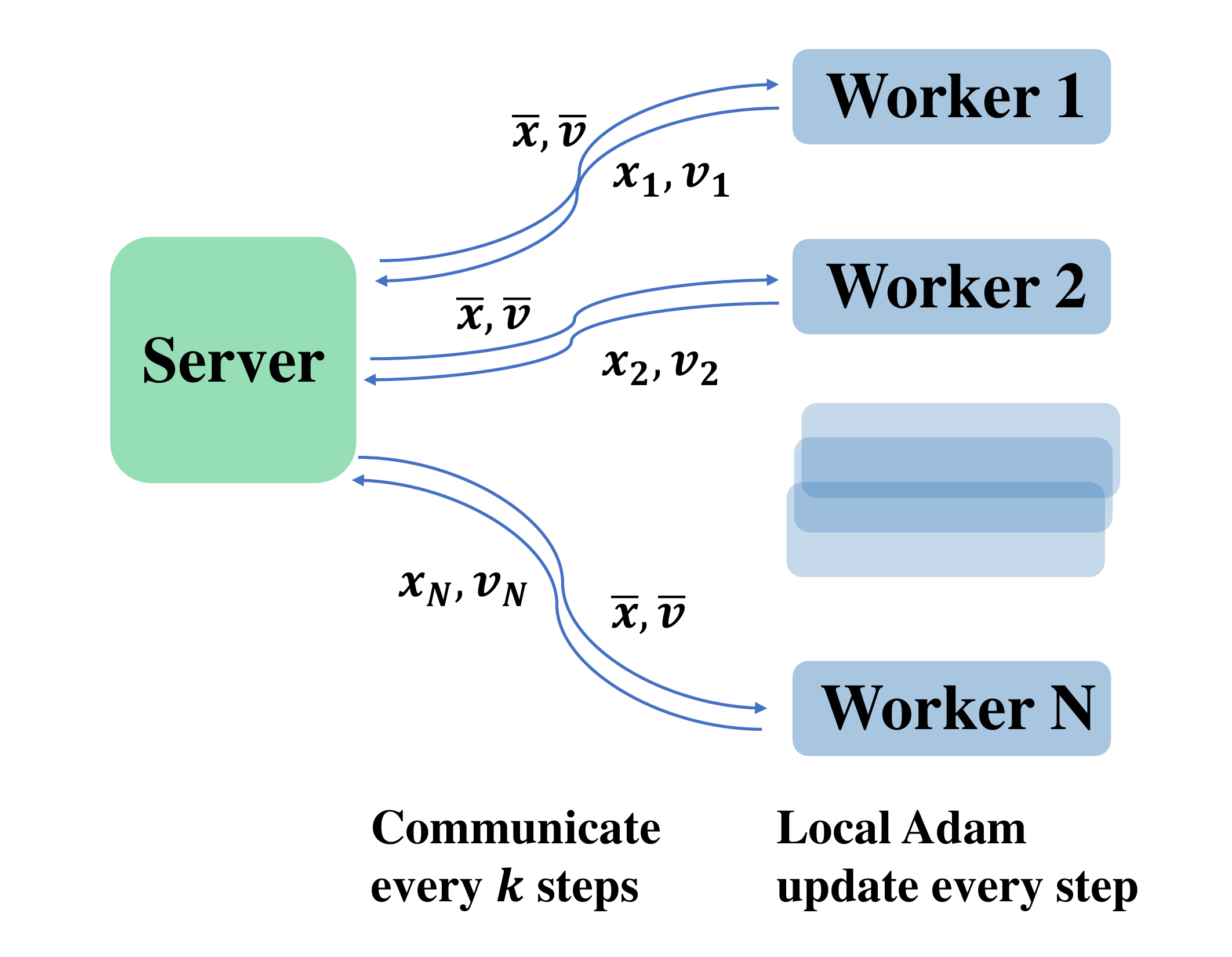}
%\vspace{-.3cm}
\caption{Illustration of distributed $k$-step Adam training scheme.}
\label{fig:k-step example}
%\vspace{-.3cm}
\end{figure}

In this section, we introduce our solution to reduce the communication cost among the computing nodes, inspired by the idea of federated learning~\citep{Proc:McMahanMRHA_AISTATS17}. Note that our CTR model uses Adam~\citep{Proc:Kingma_ICLR15} as the optimizer. In the standard distributed setting, we can average the gradients computed on different nodes in every iteration, and use the averaged gradient for an Adam update. To reduce the communication, instead of communicating in every training iteration, the ``local models'' are aggregated every $k$ step, where $k$ is a positive integer. The pseudocode is summarized in Algorithm~\ref{alg: k-step adam}, with an illustration in Figure~\ref{fig:k-step example}. On each local worker, standard Adam updates are applied. Every $k$ iterations, the local machines transmit local model parameters to the central server. The global weight is updated by taking the average of local weights, and then the new global weight is broadcast to local nodes triggering next round of local training. To ensure convergence, the second moment moving average, $v_t$, is also averaged in each communication round. To our knowledge, this is the first time a $k$-step adaptive optimization method is applied to industrial-scale massive CTR models in literature, which turns out to attain significant communication reduction with good performance (as will be shown in Section~\ref{sec:experiments}).

\vspace{0.1in}
\noindent\textbf{Communication reduction.} We highlight here the saving in communication of our approach. On a high level, the $k$-step merging method differs from standard distributed learning in the frequency of communications. In distributed learning, the server requests the \textit{local gradients} in \textbf{every iteration} to update the global model, while the communication in Algorithm~\ref{alg: k-step adam} only takes place \textbf{every $k$ iterations}, and the \textit{local models} are transmitted, which have same size as the gradients. Naturally, in principle, our method will reduce the communication cost by a factor of $1/k$. Typically, $k$ should not be too large to ensure good learning performance, since local models may deviate from the global model significantly during local training. The requirement on $k$ can be characterized by the convergence analysis, which will be presented in the remaining part of this section.

\begin{algorithm}[h]
	\caption{$k$-step Adam (with N workers)}
	\begin{algorithmic}[1]
	\STATE {\bfseries Input:} learning rate $\alpha$, $\beta_1$, $\beta_2$
	\STATE {\bfseries Initialize:} local models $x_{0,i}$, moment estimators $m_{0,i} = 0, v_{0,i} = \epsilon \mathbf{1}, i=1,...,N$
	\FOR {$t=1,...,T$}
	\STATE  Compute stochastic gradient $g_{t,i}  \leftarrow \nabla f_i(x_{t,i}) + \xi_{t,i}$
	\STATE $m_{t,i} = \beta_1 m_{t-1,i} + (1-\beta_1) g_{t,i}$
	\STATE $v_{t,i} = \beta_2 v_{t-1,i}+(1-\beta_2)g_{t,i}^2$
	\IF{$t \mod k \neq 0$}
	\STATE ${v}_{t} = {v}_{t-1}$  	
	\STATE $x_{t+1,i} \leftarrow x_{t,i} - \alpha  \frac{m_{t,i}}{\sqrt{{v}_t}}$
	\STATE {\color{blue} //Local Adam Update}
	\ELSE
	\STATE ${v}_{t} =\frac{1}{N}\sum_{i=1}^N v_{t,i} $
	\STATE $x_{t+1,i} \leftarrow \frac{1}{N}\sum_{j=1}^{N}\left(x_{t,j} - \alpha  \frac{m_{t,j}}{\sqrt{{v}_t}}\right)$
	\STATE {\color{blue} //Global averaging}
	\ENDIF
	\ENDFOR
	\end{algorithmic}
	\label{alg: k-step adam}
\end{algorithm}

In this paper, we consider the finite-sum optimization problem of the form
\begin{equation*}
    \min_x f(x)\coloneqq \frac{1}{N}\sum_{i=1}^N f_i(x),
\end{equation*}
where $N$ is the number of local computing nodes and $f_i$ is the non-convex loss function evaluated over the data on the $i$-th node. The following assumptions will be needed:

\textbf{A1:} $f_i$ is differentiable and $L$-smooth, i.e., $\|\nabla f_i(x) - \nabla f_i(y)\| \leq L\|x-y\|,\ \forall x,y$.

\textbf{A2:} Bounded and unbiased stochastic gradient, $\mathbb E [g_{t,i}] = \nabla f_i(x_{t,i})$, $\|g_{t,i}\|_{\infty} \leq G$, and bounded variance, $\mathbb E [(g_{t,i})_j^2] \leq \sigma^2, \forall j \in [d] $.

\textbf{A3:}  $\mathbb E \left[\sum_{t=1}^T\Vert \frac{1}{\sqrt{v_{t-1}}}-\frac{1}{\sqrt{v_t}} \Vert_1\right] \leq Md\cdot T^\gamma$, with some $0\leq \gamma\leq 1/2$ and $M>0$.

\vspace{0.1in}

Assumption \textbf{A1} (Lipschicity) and \textbf{A2} (bounded gradients) are standard in the optimization literature. \cite{Proc:Chen_ICLR19} has shown that the LHS term in \textbf{A3} plays an important role in the convergence of Adam-type algorithms. If it is too big, Adam may converge very slowly, or even diverge, in simple optimization problems. Thus, \textbf{A3} is to guarantee that the algorithm can converge in a proper manner. In practice, this term is usually very small. The special case $\gamma=0$ implies AMSGrad~\citep{Proc:Reddi_ICLR18,chen2020toward}, a variant of Adam known to have better theoretical convergence behavior, though they perform similarly in practice. In our analysis, we consider \textbf{A3} as a more general setting.

\begin{theorem}\label{thm: k-step adam}
Under \textbf{A1-A3}, define $\overline x_t = \frac{1}{N} \sum_{i=1} ^N x_{t,i}$ and set $\alpha = \min(\frac{\sqrt{N}}{\sqrt{Td}},\frac{\sqrt{\epsilon}}{4L})$. We have
\begin{align*}
    &\frac{1}{T}  \sum_{t=1} ^T \mathbb E \left[ \left\| \frac{ {\nabla f} (\overline x_t)}{v_t^{1/4}} \right\|^2\right] \leq  O(\frac{\sqrt d}{\sqrt TN})+O(\frac{d}{T^{1-\gamma}}+\frac{\sqrt dN}{T^{1.5-\gamma}})+O(\frac{Nk^2}{T}).
\end{align*}
\end{theorem}

% \frac{\sqrt{d}\triangle}{\sqrt{TN}}  + \frac{L\sqrt{d}}{\sqrt{TN}}  \sigma^2  \frac{1}{\epsilon}+\frac{1}{T^{1-\gamma}}\frac{\beta_1^2G^2Md}{(1-\beta_1)^2 } \nonumber\\
    % &+ \frac{\sqrt{N} L}{T^{1.5-\gamma}}  \frac{\beta_1^2G^2M\sqrt d}{\epsilon^{0.5}(1-\beta_1)^2 } +  \frac{NLG^2}{T\epsilon^{1.5}}  \left( \frac{\beta_1^2}{(1-\beta_1)^2} +5(k-1)^2  \right).

We have simplified many constants for the ease of presentation. The left-hand side of Theorem~\ref{thm: k-step adam} is a standard metric to measure the convergence speed towards a stationary point in non-convex optimization literature. The dependence on parameter dimension $d$ is due to the dimension-wise bounds in the assumptions, which can be easily removed by assuming alternative total bounds. We simplify the convergence rate in Corollary~\ref{cor: rate} when $k$ is not too large.

%\newpage

\begin{corollary}\label{cor: rate}
Under \textbf{A1-A3}, with $\alpha = \min(\frac{\sqrt{N}}{\sqrt{Td}},\frac{\sqrt{\epsilon}}{4L})$ and $k \leq O(T^{1/4}d^{1/4}/N^{3/4})$,
\begin{align}
   & \frac{1}{T}  \sum_{t=1} ^T \mathbb E \left[ \left\| { {\nabla f} (\overline x_t)} \right\|^2\right]
    \leq  O\left(\frac{1}{\sqrt{TN}}\right). \label{eqn:coro: rate}
\end{align}
\end{corollary}

\noindent\textbf{Discussion.} From Corollary~\ref{cor: rate}, we see that as long as the number of local updates $k\leq O(T^{1/4})$, the convergence rate of $k$-step Adam is $O(\frac{1}{\sqrt {TN}})$, which matches that of $k$-step SGD (FedAvg/LocalSGD) e.g.,~\cite{zhou2017convergence,Porc:Yu_ICML19}, and achieves linear speedup compared with vanilla Adam-type method~\citep{Arxiv:Zhou_18,Proc:Chen_ICLR19} w.r.t. $N$. For the left-hand side of (\ref{eqn:coro: rate}) to achieve $O(\delta)$, a single worker needs $O(1/\epsilon^2)$ iterations, while using $N$ workers as in $k$-step Adam only requires $O(1/N\epsilon^2)$ iterations, leading to a linear acceleration in training efficiency (in wall-clock time). The condition on $k$ suggests that, we can attain the same convergence rate of vanilla Adam, with the number of communication rounds sub-linear in $T$. Since in our framework the data on each computing node are identically distributed, the $k$-step averaging method is expected to work very well empirically, which is a well recognized behavior of federated learning. In the following, we provide a sketch proof of Theorem~\ref{thm: k-step adam}. While the proof is for a new algorithm, we follow the standard procedure in the literature such as~\cite{Arxiv:Zhou_18,Proc:Chen_ICLR19,chen2020toward}.

\vspace{0.4in}
\allowdisplaybreaks
\noindent\textbf{Sketch Proof of Theorem~\ref{thm: k-step adam}:}
 Define an auxiliary sequence
\begin{align*} %\label{eq: def_z}
 \overline{\theta}_t = \overline{x}_t + \frac{\beta_1}{1-\beta_1}(\overline x_t - \overline x_{t-1} ).
\end{align*}
where $\overline x_t = \frac{1}{N} \sum_{i=1} ^N x_{t,i}$ and let $x_0 \triangleq x_1$. It can be shown that
\begin{equation}\label{eq: z_update}
    \overline \theta_{t+1} - \overline \theta_{t} = \alpha \frac{\beta_1}{1-\beta_1 } \left(\frac{1}{\sqrt{v_{t-1}}} - \frac{1}{\sqrt{v_{t}}}\right) \odot \overline m_{t-1} - \alpha \frac{\overline g_t}{\sqrt{v_t}},
\end{equation}
where $\overline m_{t} = \frac{1}{N}  \sum_{i=1}^ N m_{t,i}$ and  $\overline g_{t} = \frac{1}{N} \sum_{i=1}^ N g_{t,i}$. By \textbf{A1},
\begin{align*}
    f(\overline  \theta_{t+1}) \leq     f(\overline  \theta_{t}) + \langle \nabla f(\overline \theta_t ), \overline \theta_{t+1} - \overline \theta_t \rangle + \frac{L}{2 } \|\overline \theta_{t+1} - \overline \theta_t \|^2. \nonumber
\end{align*}
Hence,
\begin{align}
      &-\mathbb E [ \langle \nabla f(\overline \theta_t ), \overline \theta_{t+1} - \overline \theta_t \rangle] \leq   \mathbb E[   f(\overline  \theta_{t})-f(\overline  \theta_{t+1})] + \frac{L}{2 }\mathbb E[ \|\overline \theta_{t+1} - \overline \theta_t \|^2]. \label{eq: lipschitz_exp}
\end{align}
where the expectation is taken over all the randomness of stochastic gradients until iteration $t$. By~\eqref{eq: z_update}, we have
\begin{align}\label{eq: first_order}
     \langle \nabla f(\overline \theta_t ), \overline \theta_{t+1} - \overline \theta_t \rangle
    =   \alpha \langle  \nabla f(\theta_t), \frac{\beta_1}{1-\beta_1 } \left(\frac{1}{\sqrt{v_{t-1}}} - \frac{1}{\sqrt{v_{t}}}\right) \odot \overline m_{t-1}
     - \alpha \langle \nabla f(\theta_t), \frac{\overline g_t}{\sqrt{v_t}} \rangle.
\end{align}
Since $v_{t}$ is independent of $\overline g_t$ and $\mathbb E[g_{t,i}] = \nabla f_i(x_{t,i})$, taking expectation on both sides of~\eqref{eq: first_order} yields
\begin{align*}
     \mathbb E \left[ \langle \nabla f(\overline \theta_t ), \overline \theta_{t+1} - \overline \theta_t \rangle \right]
    =     \alpha \mathbb E \left [ \langle  \nabla f(\theta_t), \frac{\beta_1}{1-\beta_1 } \left(\frac{1}{\sqrt{v_{t-1}}} - \frac{1}{\sqrt{v_{t}}}\right) \odot \overline m_{t-1}\right ]
     - \alpha \mathbb E \big[ \langle \nabla f(\theta_t), \frac{ \overline {\nabla f} (x_t)}{\sqrt{v_t}} \rangle  \big]. \nonumber
\end{align*}
where $\overline {\nabla f} (x_t) = \frac{1}{N}\sum_{i=1}^N \nabla f_i (x_{t,i})$. Denoting the inner product in the second expectation as $B_1$, we have
\begin{align}\label{eq: bias_split}
    B_1
    =  \frac{1}{2}\Bigg(\left \| \frac{\nabla f(\overline \theta_t)}{v_t^{1/4}}\right\|^2 + \left\| \frac{\overline {\nabla f} (x_t)}{v_t^{1/4}} \right\|^2 - \left \| \frac{\nabla f(\overline \theta_t) - \overline {\nabla f} (x_t)}{v_t^{1/4}} \right\|^2\Bigg).
\end{align}
For the last term, by Cauchy-Schwartz inequality,
\begin{align} \label{eq: bias_bound}
    \left \| \frac{\nabla f(\overline \theta_t) - \overline {\nabla f} (x_t)}{v_t^{1/4}} \right\|^2
    = & \left \| \frac{\frac{1}{N} \sum_{i=1}^N ( \nabla f_i(\overline \theta_t) -  {\nabla f_i} (x_{t,i}))}{v_t^{1/4}} \right\|^2  \nonumber \\
    \leq & \frac{2}{N} \sum_{i=1} ^ N  \left(\Bigg \| \frac{\nabla f_i(\overline \theta_t) -  \nabla f_i(\overline x_{t})}{v_t^{1/4}} \right\|^2 + \left \| \frac{\nabla f_i(\overline x_t) -  {\nabla f_i} (x_{t,i})}{v_t^{1/4}} \right\|^2 \Bigg) .
\end{align}
Using Lipschitz property (Assumption A1) of $\nabla f_i$, we can further bound the differences of gradients on RHS of~\eqref{eq: bias_bound} by
\begin{align}
   &\frac{2}{N} \sum_{i=1} ^ N    \left \| \frac{\nabla f_i(\overline \theta_t) - \nabla f_i(\overline x_{t})}{v_t^{1/4}} \right\|^2
    \leq  \frac{2}{N} \sum_{i=1} ^ N \frac{L \|\overline \theta_t -   \overline x_{t}\|^2  }{\epsilon^{0.5}} , \label{eq: z_x_lip} \\
       & \frac{2}{N} \sum_{i=1} ^ N \left \| \frac{\nabla f_i(\overline x_t) -  {\nabla f_i} (x_{t,i})}{v_t^{1/4}} \right\|^2
    \leq \frac{2}{N} \sum_{i=1} ^ N  \frac{L\| \overline x_{t} - x_{t,i}\|^2 }{\epsilon^{0.5}} . \label{eq: x_x_avg_diff}
\end{align}

It remains to bound $\|\overline \theta_t - \overline x_t\|^2$ and $\|\overline x_t - x_{t,i}\|^2$ using the update rule of $x$ and $\theta$.
%First, we can split the different into two parts as
% \begin{align}
% \overline \theta_t - x_{t,i} =& \overline \theta_t - \overline x_t + \overline x_t - x_{t,i} \nonumber \\
% =& \frac{\beta_1}{1-\beta_1} (\overline x_t - \overline x_{t-1}) + \overline x_t - x_{t,i} \nonumber
% % \end{align}
% where the first part is iteration difference and the second part is consensus error.
For the difference between $\overline \theta_t$ and $\overline x_t$, we have
% \begin{align}\label{eq: z_x_i}
%      \sum_{i=1} ^ N  \left \| {\overline \theta_t -   x_{t,i}}{} \right\|^2 \leq &   2 \frac{\beta_1^2}{(1-\beta_1)^2} \sum_{i=1} ^ N   \left \| {\overline x_t - \overline x_{t-1}}{} \right\|^2 \nonumber \\
%      &+ \sum_{i=1} ^ N 2 \left \| {\overline x_t - x_{t,i}}{} \right\|^2
% \end{align}
% Now we need to bound the two terms on RHS of~\eqref{eq: z_x_i}. For the first term, we have
\begin{align} \label{eq: z_x_bound}
\sum_{i=1} ^ N  \left \| {\overline \theta_t -   \overline x_{t}} \right\|^2  = &\frac{\beta_1^2}{(1-\beta_1)^2} \sum_{i=1} ^ N   \left \| {\overline x_t - \overline x_{t-1}}{} \right\|^2  \nonumber \\
= & \frac{\beta_1^2}{(1-\beta_1)^2}   \alpha^2  N\left\| \frac{\overline m_{t-1}}{ \sqrt{v_{t-1}}}\right\|^2
\leq \frac{\beta_1^2}{(1-\beta_1)^2} \alpha^2 Nd  \frac{G^2}{\epsilon}.
\end{align}

For the consensus error~\eqref{eq: x_x_avg_diff}, let $\floor{t}_k$ be the largest multiple of $k$ that is less than $t$. By the updating rule, we have
\begin{align}
     &\sum_{i}^N\left \| {\overline x_t - x_{t,i}} \right\|^2
     =  \alpha ^2 \sum_{i=1}^N \left \|   \sum_{l = \floor{t}_k +1 }^{t-1} \left( \frac{m_{l,i}}{\sqrt{v_l}}  - \frac{1}{N}\sum_{o=1} ^ N \frac{m_{l,o}}{\sqrt{v_l}}  \right) \right\|^2
    \leq  4 N(k-1)^2 \alpha^2 d \frac{G^2}{\epsilon}, \label{eq: consensus_bound}
\end{align}
since $(v_o)_j \geq \epsilon,  \forall j \in [d], o \in [N]$, and $\|m_{t,o}\|_{\infty} \leq G, \forall t, o$. Combining~\eqref{eq: bias_bound},~\eqref{eq: z_x_lip},~\eqref{eq: x_x_avg_diff},~\eqref{eq: z_x_bound} and~\eqref{eq: consensus_bound} gives
\begin{align}\label{eq: bias_overall_bound}
   &\left \| \frac{\nabla f(\overline \theta_t) - \overline {\nabla f} (x_t)}{v_t^{1/4}} \right\|^2
    \leq  \frac{\alpha^2 L d G^2}{\epsilon^{1.5}} \left( \frac{2\beta_1^2}{(1-\beta_1)^2} +8(k-1)^2  \right) .
\end{align}
Now, combining~\eqref{eq: bias_split}, \eqref{eq: bias_overall_bound}, \eqref{eq: first_order} and~\eqref{eq: lipschitz_exp}, telescoping summation gives
% \begin{align*}
%     &\alpha \mathbb E \left[ \frac{1}{2}\left \| \frac{\nabla f(\overline \theta_t)}{v_t^{1/4}}\right\|^2 + \frac{1}{2}\left\| \frac{\overline {\nabla f} (x_t)}{v_t^{1/4}} \right\|^2\right]
%     \\
%     \leq&   \mathbb E[   f(\overline  \theta_{t})] -  \mathbb E [f(\overline  \theta_{t+1})] + \frac{L}{2 }\mathbb E[ \|\overline \theta_{t+1} -  \overline \theta_t \|^2] \\
%     +& \alpha \mathbb E \left [ \langle  \nabla f(\overline \theta_t), \frac{\beta_1}{1-\beta_1 } \left(\frac{1}{\sqrt{v_{t-1}}} - \frac{1}{\sqrt{v_{t}}}\right) \odot \overline m_{t-1}\right ]  \nonumber \\
%     +& \alpha^3  \frac{LdG^2}{\epsilon^{1.5}} \left( \frac{\beta_1^2}{(1-\beta_1)^2} +4(k-1)^2  \right). \nonumber
% \end{align*}
\begin{align}\label{eq: remain_t1_t2}
    \frac{1}{T}  \sum_{t=1} ^T \mathbb E \left[ \frac{1}{2}\left \| \frac{\nabla f(\overline \theta_t)}{v_t^{1/4}}\right\|^2 + \frac{1}{2}\left\| \frac{\overline {\nabla f} (x_t)}{v_t^{1/4}} \right\|^2\right]
    \leq  & \frac{\mathbb E[   f(\overline  z_{1})] -  \mathbb E [f(\overline  z_{T+1})]}{T\alpha} + \frac{L}{2 } \frac{1}{T\alpha} \underbrace{\sum_{t=1}^T \mathbb E[ \|\overline \theta_{t+1} - \overline \theta_t \|^2]}_{A_1}  \nonumber \\
    &\hspace{-0.2in}+ \frac{1}{T}\underbrace{\sum_{t=1} ^T \mathbb E \left [ \left \langle  \nabla f(\overline \theta_t), \frac{\beta_1}{1-\beta_1 } \left(\frac{1}{\sqrt{v_{t-1}}} - \frac{1}{\sqrt{v_{t}}}\right) \odot \overline m_{t-1}\right\rangle \right ]}_{A_2}  \nonumber \\
    &+  \alpha^2 \frac{LdG^2}{\epsilon^{1.5}} \left( \frac{\beta_1^2}{(1-\beta_1)^2} +4(k-1)^2  \right).
\end{align}
To bound $A_1$, by~\eqref{eq: z_update}, we know that
%\newpage
\begin{align*}
    \|\overline \theta_{t+1} - \overline \theta_t \|^2
    \leq   2\alpha^2 \left(\left \| \frac{\beta_1}{1-\beta_1 } \left(\frac{1}{\sqrt{v_{t-1}}} - \frac{1}{\sqrt{v_{t}}}\right) \odot \overline m_{t-1} \right\|^2 +  \left\|  \frac{\overline g_t}{\sqrt{v_t}}\right\|^2\right). \nonumber
\end{align*}
In addition, conditioned on the $\sigma$-field $\mathcal F_{t-1}$ (until $t-1$), we have
\begin{align*}
    \mathbb E \left[  \left\|  \frac{\overline g_t}{\sqrt{v_t}}\right\|^2\right]
    =& \frac{1}{N^2}  \mathbb E \left[ \sum_{j=1}^N \sum_{i=1}^N  \left \langle \frac{ g_{t,i}}{\sqrt{v_t}},\frac{ g_{t,j}}{\sqrt{v_t}} \right\rangle \right]  \nonumber \\
    \stackrel{(a)}{=} &  \frac{1}{N^2}  \mathbb E \left[ \sum_{j=1}^N \sum_{i=1}^N  \left \langle \frac{ \nabla f_i(x_{t,i}) + \xi_{t,i}}{\sqrt{v_t}},\frac{\nabla f_j(x_{t,j}) + \xi_{t,j}}{\sqrt{v_t}} \right\rangle \right] \nonumber \\
    \stackrel{(b)}{=} &  \frac{1}{N^2} \left\| \frac{ \sum_{i=1}^N \nabla f_i(x_{t,i})}{\sqrt{v_t}}\right\|^2 +  \frac{1}{N^2} \sum_{i=1}^N   \mathbb E \left[ \left\| \frac{ \xi_{t,i}}{\sqrt{v_t}} \right\|^2  \right]
    \stackrel{(c)}{\leq} \left\| \frac{\overline {\nabla f}  (x_{t})}{\sqrt{v_t}}\right\|^2 +  \frac{1}{N\epsilon}   \sigma^2. \nonumber
\end{align*}
where $(a)$ is a decomposition of stochastic gradient, and $(b)$ and $(c)$ are due to \textbf{A2}. Furthermore, by \textbf{A3} we have
\begin{align*}
    \mathbb E \left[\sum_{t=1}^T \left \| \frac{\beta_1}{1-\beta_1 } \left(\frac{1}{\sqrt{v_{t-1}}} - \frac{1}{\sqrt{v_{t}}}\right) \odot \overline m_{t-1} \right\|^2\right]
    \leq & \mathbb E \left[\sum_{t=1}^T \frac{\beta_1^2}{(1-\beta_1)^2 } G^2 \frac{1}{\epsilon^{0.5}}  \sum_{j=1}^d
\left|\frac{1}{\sqrt{(v_{t-1})_j}} - \frac{1}{\sqrt{(v_{t})_j}} \right|\right] \nonumber \\
\leq & \frac{\beta_1^2G^2MdT^\gamma}{\epsilon^{0.5}(1-\beta_1)^2 }.   \nonumber
\end{align*}
Combining above results, we obtain
\begin{align*}\notag
    A_1 \leq & 2\alpha^2 \sum_{t=1}^T \mathbb E \left [\left\| \frac{\overline {\nabla f}  (x_{t})}{\sqrt{v_t}}\right\|^2 +  \frac{1}{N\epsilon}\sigma^2    \right]
    +  2\alpha^2  \frac{\beta_1^2G^2MdT^\gamma}{\epsilon^{0.5}(1-\beta_1)^2 }.
\end{align*}
We can use similar argument to get $A_2  \leq \frac{\beta_1^2G^2MT^\gamma}{(1-\beta_1)^2 }$. Returning to~\eqref{eq: remain_t1_t2}, since $z_1 = x_1$, we have
\begin{align*}%\label{eq: semi_final}
   \frac{1}{T}  \sum_{t=1} ^T \mathbb E \left[ \frac{1}{2}\left \| \frac{\nabla f(\overline \theta_t)}{v_t^{1/4}}\right\|^2 + \frac{1}{2}\left\| \frac{\overline {\nabla f} (x_t)}{v_t^{1/4}} \right\|^2\right]
    \leq  & \frac{\mathbb E[   f(\overline  x_{1})] -  \mathbb E [f(\overline  z_{T+1})]}{T\alpha} +\frac{L}{T} \alpha \sum_{t=1}^T \mathbb E \left [\Bigg\| \frac{\overline {\nabla f}  (x_{t})}{\sqrt{v_t}}\right\|^2 +  \frac{1}{N\epsilon} \sigma^2 \Bigg] \\
    & \hspace{-1.3in}+ \frac{1}{T^{1-\gamma}}\frac{\beta_1^2G^2Md}{(1-\beta_1)^2 } + \frac{\alpha L}{T^{1-\gamma}}  \frac{\beta_1^2G^2Md}{\epsilon^{0.5}(1-\beta_1)^2 } +  \alpha^2 \frac{LdG^2}{\epsilon^{1.5}} \left( \frac{\beta_1^2}{(1-\beta_1)^2} +4(k-1)^2  \right).
\end{align*}
By choosing $\alpha = \min(\frac{\sqrt{N}}{\sqrt{Td}},\frac{\sqrt{\epsilon}}{4L})$, we have
\begin{align*} %\label{eq: cancel_descent}
    \frac{L}{T} \alpha \sum_{t=1}^T \mathbb E \left [\left\| \frac{\overline {\nabla f}  (x_{t})}{\sqrt{v_t}}\right\|^2  \right]
    \leq & \frac{1}{4} \frac{1}{T}  \sum_{t=1}^T \mathbb E \left [\left\| \frac{\overline {\nabla f}  (x_{t})}{{v_t^{1/4}}}\right\|^2  \right].
\end{align*}

\newpage

Let $\triangle=\mathbb E[   f(\overline  x_{1})] -  \min_x f(x)$, we obtain
\begin{align*}
   \frac{1}{T}  \sum_{t=1} ^T \mathbb E \left[ \frac{1}{2}\left \| \frac{\nabla f(\overline \theta_t)}{v_t^{1/4}}\right\|^2 + \frac{1}{4}\left\| \frac{\overline {\nabla f} (x_t)}{v_t^{1/4}} \right\|^2\right]
    \leq  & \frac{\sqrt{d}\triangle}{\sqrt{TN}}
    + \frac{L\sqrt{d}}{\sqrt{TN}}  \sigma^2  \frac{1}{\epsilon}  +\frac{1}{T^{1-\gamma}}\frac{\beta_1^2G^2Md}{(1-\beta_1)^2 }\nonumber \\
    &\hspace{-0.3in} + \frac{\sqrt{N} L}{T^{1.5-\gamma}}  \frac{\beta_1^2G^2M\sqrt d}{\epsilon^{0.5}(1-\beta_1)^2 }  +  \frac{NLG^2}{T\epsilon^{1.5}}  \left( \frac{\beta_1^2}{(1-\beta_1)^2} +4(k-1)^2  \right).  \nonumber
\end{align*}
Note that by C-S and Jensen's inequality,
\begin{align*} %\label{eq: descent_split}
     \left\| \frac{\overline {\nabla f} (x_t)}{v_t^{1/4}} \right\|^2
     \geq &  \frac{1}{2}\left\| \frac{ {\nabla f} (\overline x_t)}{v_t^{1/4}} \right\|^2- \left\| \frac{\overline {\nabla f} (x_t) - {\nabla f} (\overline x_t)}{v_t^{1/4}} \right\|^2 \nonumber \\
     \geq &  \frac{1}{2}\left\| \frac{ {\nabla f} (\overline x_t)}{v_t^{1/4}} \right\|^2 -  \frac{1}{N} \sum_{i=1}^N \left\|  \frac{ {\nabla f_i} (x_{t,i}) - {\nabla f_i} (\overline x_t)}{v_t^{1/4}} \right\|^2 \nonumber \\
     \geq & \frac{1}{2}\left\| \frac{ {\nabla f} (\overline x_t)}{v_t^{1/4}} \right\|^2 - 4L(k-1)^2 \alpha^2 d \frac{G^2}{\epsilon^{1.5}}.  \nonumber
\end{align*}
The desired result then follows. This completes the proof.
\hfill $\square$

\section{Experimental Evaluation}  \label{sec:experiments}

The objective of the experimental evaluation is to investigate the performance of our system and the impact of optimizations: %Specifically, the experiments are designed to answer the following questions:
\begin{itemize}
\item What are the impacts of our proposed optimizations, including two-phase GPU communication, GPUDirect RDMA, SSD direct I/O, and core binding?
\item Does $k$-step model merging hurt the prediction accuracy?
\item How much data communication time is saved?
\end{itemize}

\vspace{0.1in}

\noindent\textbf{System.} We perform the experimental evaluation on 8 GPU computing nodes. Each node has 8 cutting-edge 32 GB HBM GPUs, server-grade CPUs with 48 cores (96 threads), $\sim$1 TB of memory, $\sim$20 TB RAID-0 NVMe SSDs and a 100 Gb RDMA network adaptor. Thus, in our setting, using $n$ nodes corresponds to $8\times k$ local distributed workers. All nodes are inter-connected through a high-speed Ethernet switch.
We use a web search CTR prediction model in real-world online sponsor advertising applications to investigate the
effectiveness of our proposed system. The number of sparse parameters is $\sim$$10^{11}$. The number of dense parameters is $\sim$$10^{6}$. The size of the model is $\sim$10 TB. For dense parameters, we use Adam~\citep{Proc:Kingma_ICLR15} optimizer. For sparse parameters, we use AdaGrad optimizer~\citep{adagrad} to avoid storing the extra first-order momentum which would take substantial space for the huge sparse layers. Since each training input only leads to a small number of non-zero gradients in the sparse layers, we adopt a standard distributed setting on sparse model parameters, where the gradients are averaged for update in every iteration. This is cheap in communication.
For dense model that takes more communication, we use $k$-step Adam as optimizer, with default parameter setting $\beta_1=0.0$, $\beta_2=0.999$.

%\vspace{0.1in}

\newpage

\noindent\textbf{Data.} We collect user click history logs from a real-world search engine in 24 hours as the training and testing dataset. The dataset is chunked into batches---each batch contains $\sim$$4$$\times$$10^{6}$ instances. The worker further splits each batch into mini-batches with $\sim$$1,000$ instances as a training step.
Note that our ``$k$-step'' is worked on a mini-batch level: the model merging is performed after every $k$ mini-batches. The training and testing are performed in an online learning fashion: an instance is first predicted using the inference result of the current model to generate the test accuracy, and then it is fed to the model for the training. The online training is ``hot-started'': we use the trained model on previous days at the start point in the experiments instead of a random initialized one.

\subsection{Core Binding and SSD Direct I/O}

\begin{figure}[h]
%\hspace{-0.2in}
\centering
\mbox{
	%\includegraphics[width=2.2in]{arxiv_fig/time_split_nodio.pdf}
	%\hspace{0.15in}
	\includegraphics[width=2.2in]{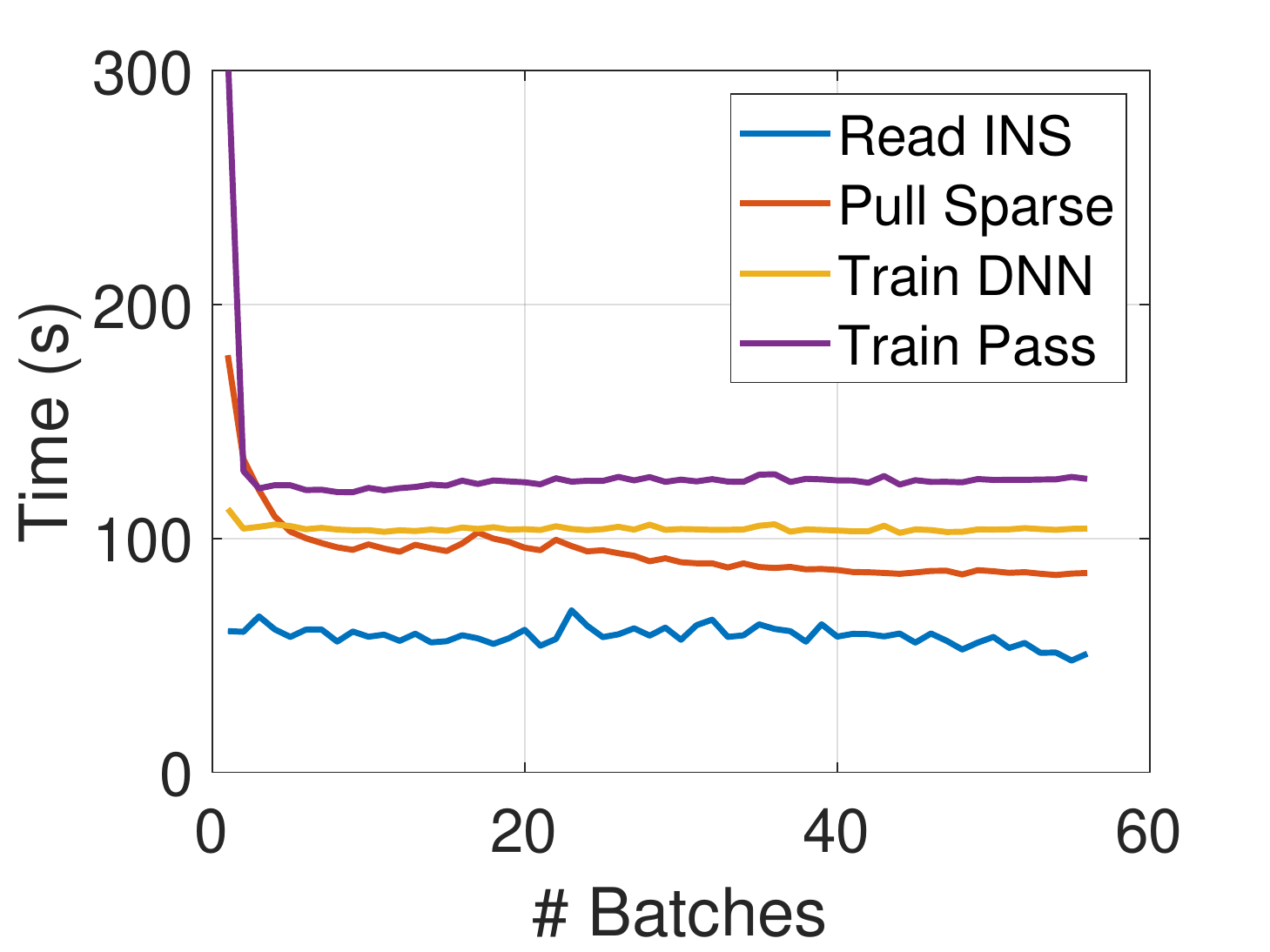}
	\hspace{0.15in}
	\includegraphics[width=2.2in]{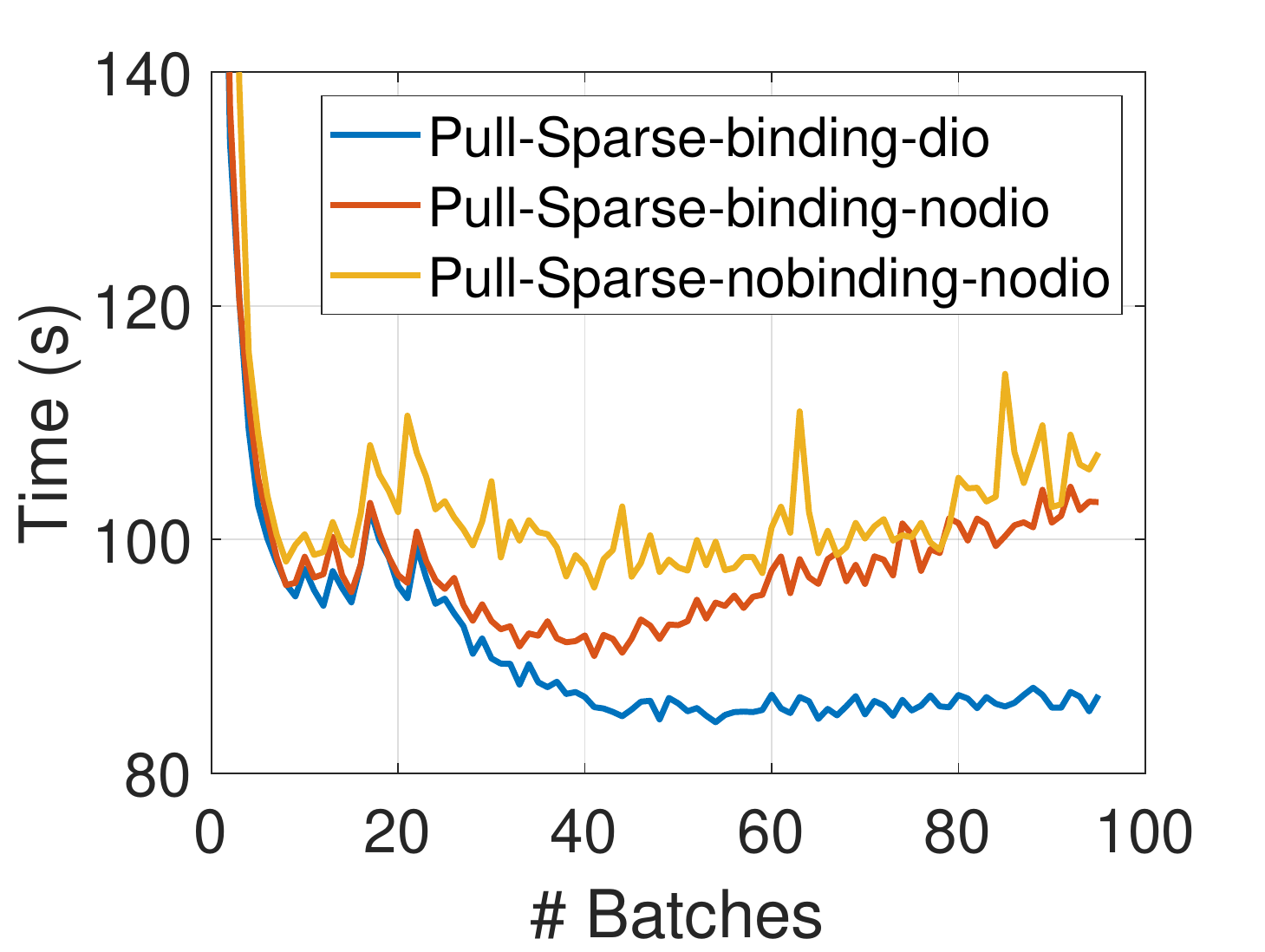}
}
%\vspace{-0.2in}
\caption{Left: Time distribution of CTR model training with all optimizations; Right: Time of pull sparse parameters with/without core binding and direct I/O.}
\label{fig:binding-dio}
\end{figure}

Figure~\ref{fig:binding-dio} illustrates the effect of using core binding and direct I/O on SSDs. The experiments are executed in one single node to eliminate the impact of the network communications. In the left part of the figure, we present the training time with core binding and direct I/O for each stage, where \texttt{Read Ins} is the time to read and parse training instances from the network file system; \texttt{Pull Sparse} is the time to load sparse parameters from main memory and SSDs to GPUs; \texttt{Train DNN} is the GPU computation time to perform feed forward and backward propagation; \texttt{Overall} is the overall execution time to train a batch. The overall time is not the summation of all three stages (Read Ins, Pull Sparse, and Train DNN) because the three stages are executed in a pipeline---the execution time of these stages are overlapped. With core binding and direct I/O, the sparse parameter loading becomes faster than the DNN training and no longer a bottleneck.
The right part of Figure~\ref{fig:binding-dio} presents the finer granularity comparison for core binding and direct I/O. These two techniques mainly optimize the \texttt{Pull Sparse} stage. Thus, we compare the execution time of pulling sparse parameters by disabling them. \texttt{Pull-binding-dio} is the configuration enabling both optimizations. Note that \texttt{Pull-binding-dio} is the same curve as the \texttt{Pull-Sparse} in the left part of the figure. The range of the y axis is magnified and more batch training is presented to show the detailed optimization impact of binding and direct I/O. \texttt{Pull-binding-nodio} and \texttt{Pull-nobinding-nodio} illustrate the configuration when we disable direct I/O and disable both core binding and direct I/O, respectively. The core binding provides $5-10\%$ improvement to the baseline scenario. Adding the direct I/O on SSDs shows another $10-20\%$ improvement. The in-memory cache management maintains the frequently used parameters well during the training progress. The direct I/O improvement becomes larger in the late stage of the training because it eliminates the more unnecessary operating system level page cache I/Os.

\subsection{GPU Communications}

\begin{figure}[htbp]
\vspace{-0.1in}
%\hspace{-0.1in}
\centering
\mbox{
	\includegraphics[width=2.2in]{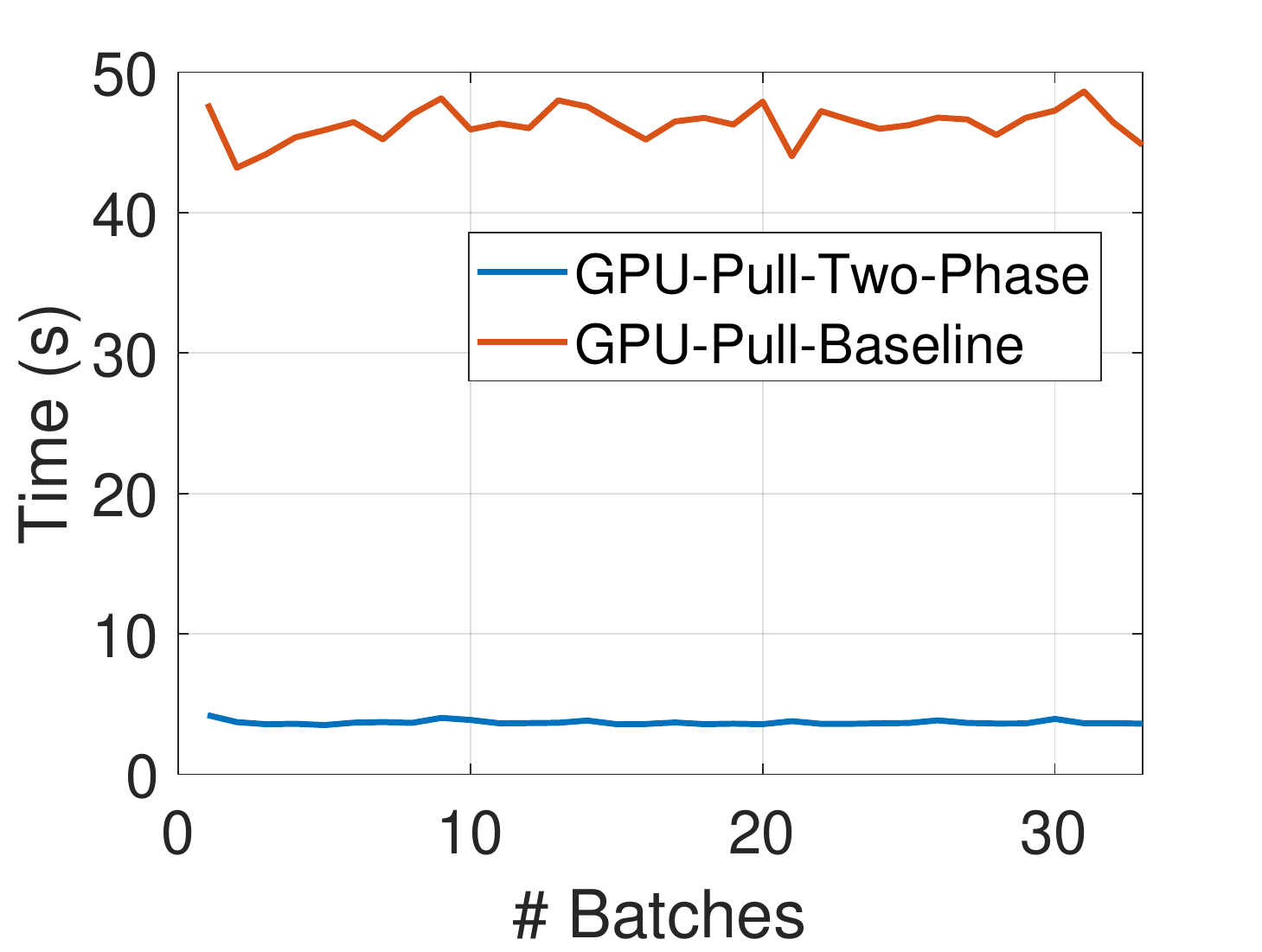}
	\hspace{0.15in}
	\includegraphics[width=2.2in]{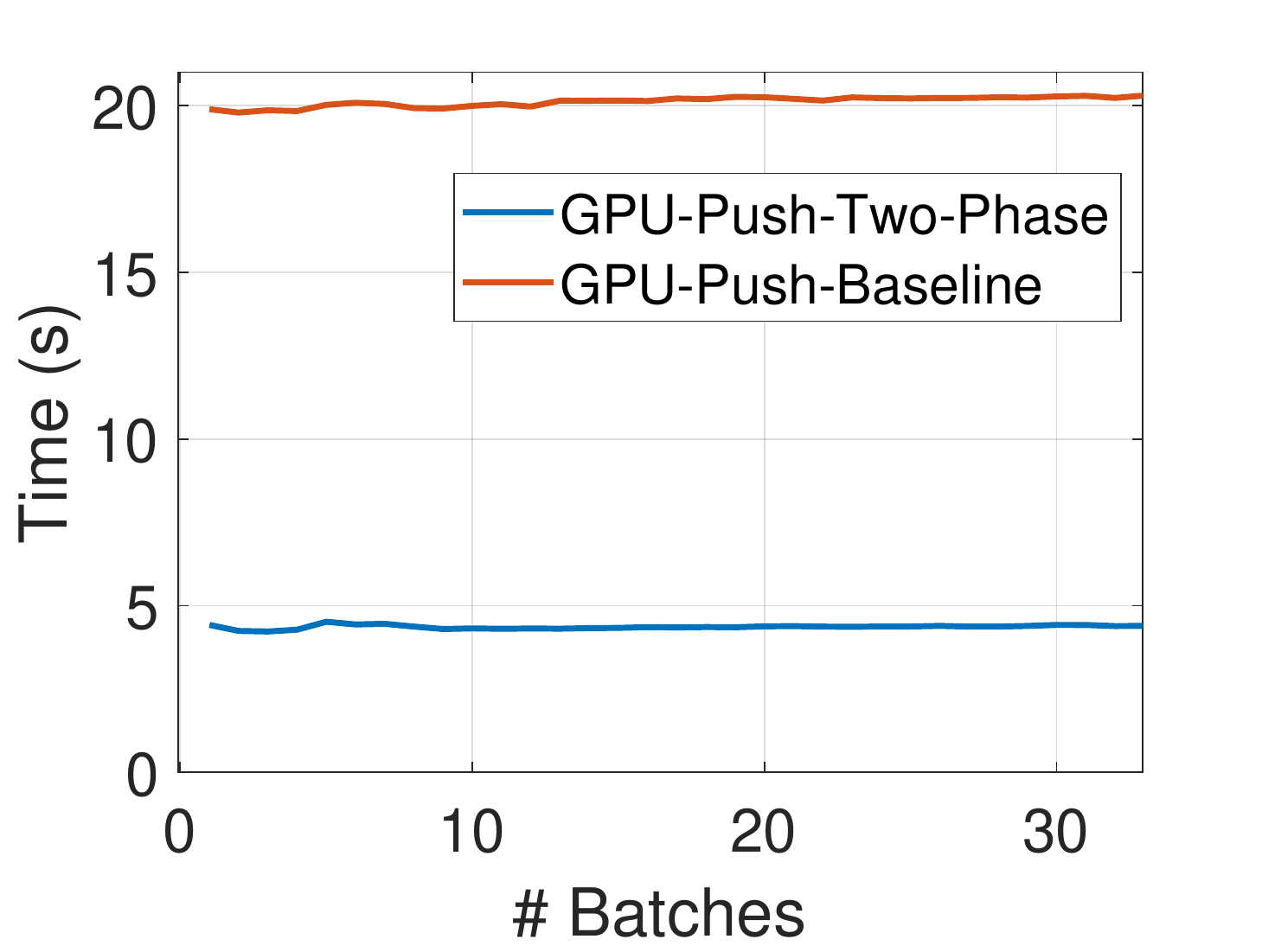}
}
\vspace{-0.15in}
\caption{Time of pull/push operations in GPU HBM with/without two-phase GPU communication.}
\label{fig:two-phase}
\end{figure}

\noindent\textbf{Two-Phase GPU Communication.}
The two-phase GPU communication technique uses a middleman GPU buffer to eliminate the data transfer on PCIe and main memory. Figure~\ref{fig:two-phase} presents the GPU communications within a node to pull the parameters in the distributed GPU hash table from the HBM of peer GPUs (\texttt{GPU-Pull}); and the time to update the parameters in the distributed GPU hash table (\texttt{GPU-Push}). As depicted in Figure~\ref{fig:two-phase}, the two-phase optimization reduces as much as $90\%$ communication time among GPUs in the same node for pulling parameters and $75\%$ improvement for pushing updates to other peer GPUs.

\begin{figure}[hp]
\vspace{-0.1in}
\centering
\mbox{
	\includegraphics[width=2.2in]{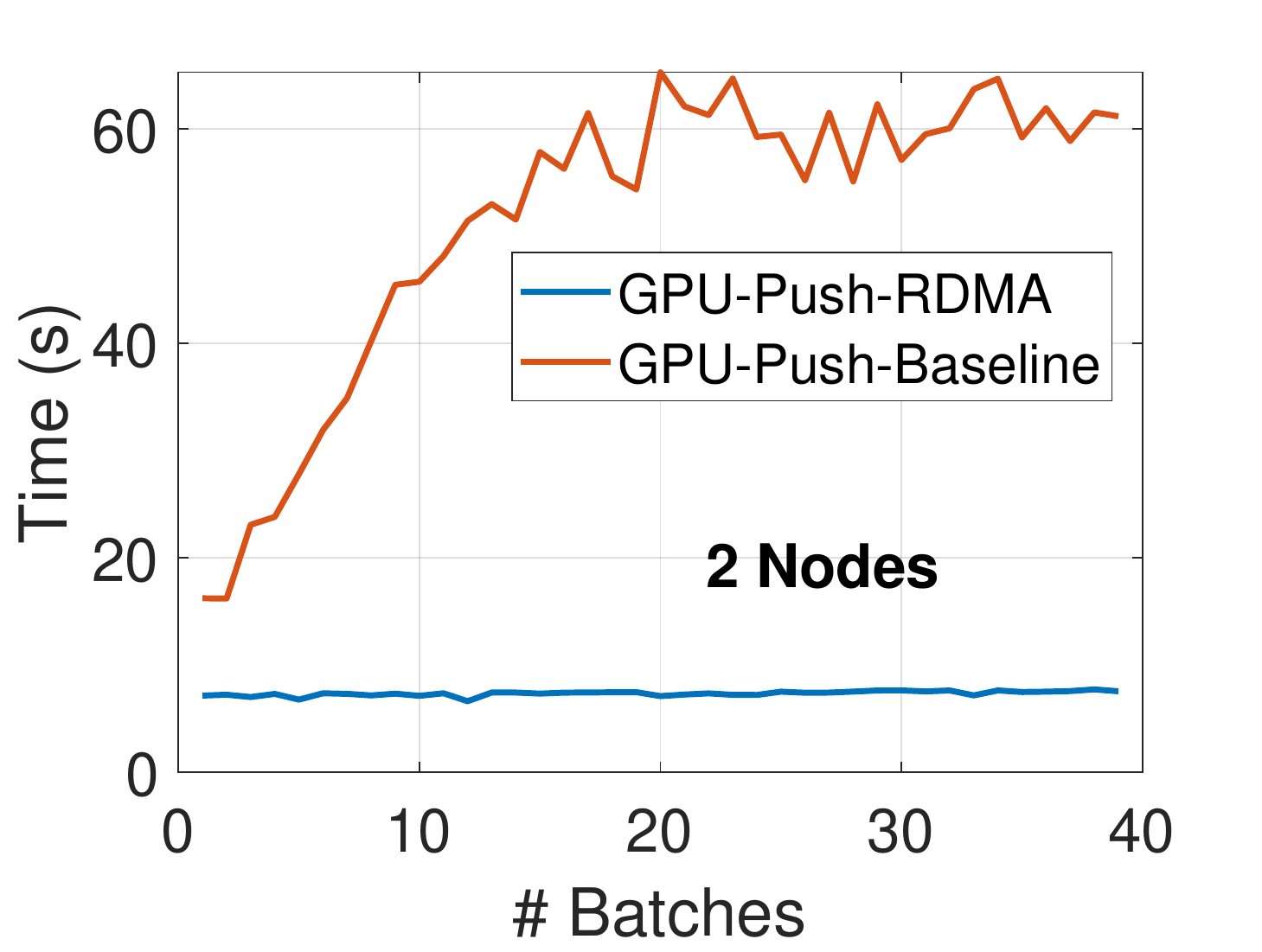}
	\hspace{0.15in}
	\includegraphics[width=2.2in]{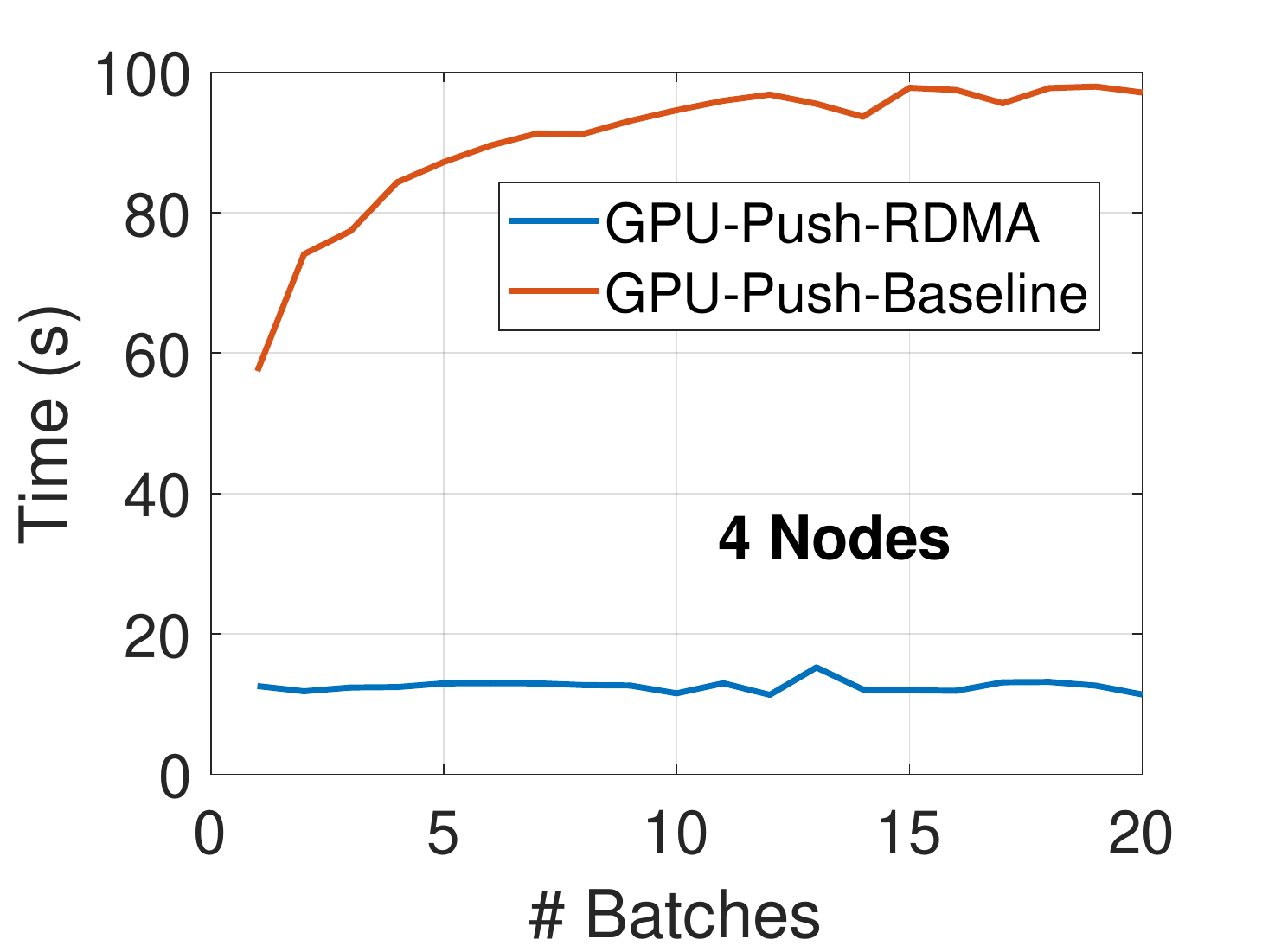}
}

\vspace{-0.15in}
\caption{Time of push operations with/without RDMA.}
\label{fig:rdma}
\end{figure}

\begin{figure}[b!]
\vspace{-0.15in}
\centering
\mbox{
	\includegraphics[width=2.2in]{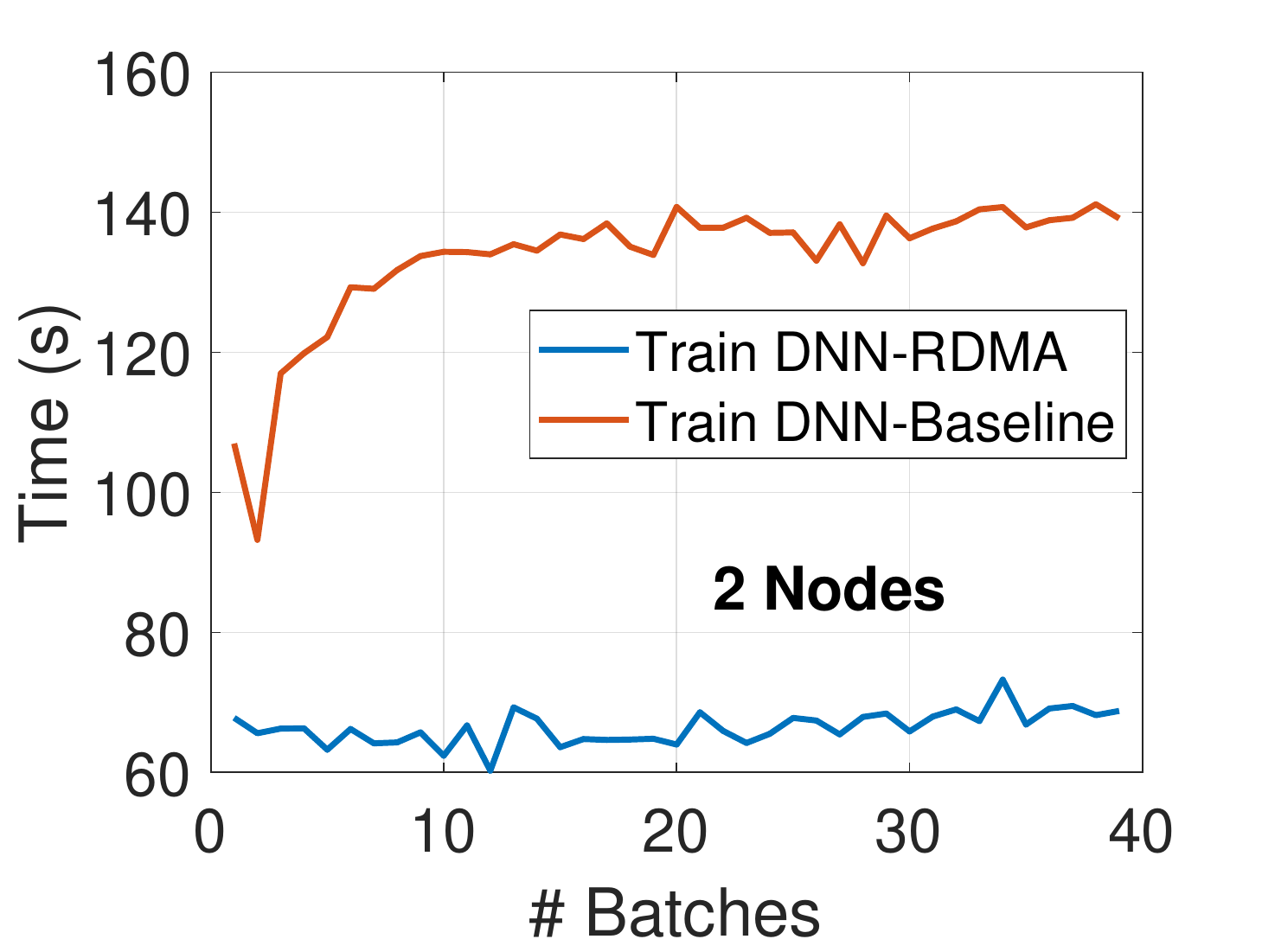}
	\hspace{0.15in}
	\includegraphics[width=2.2in]{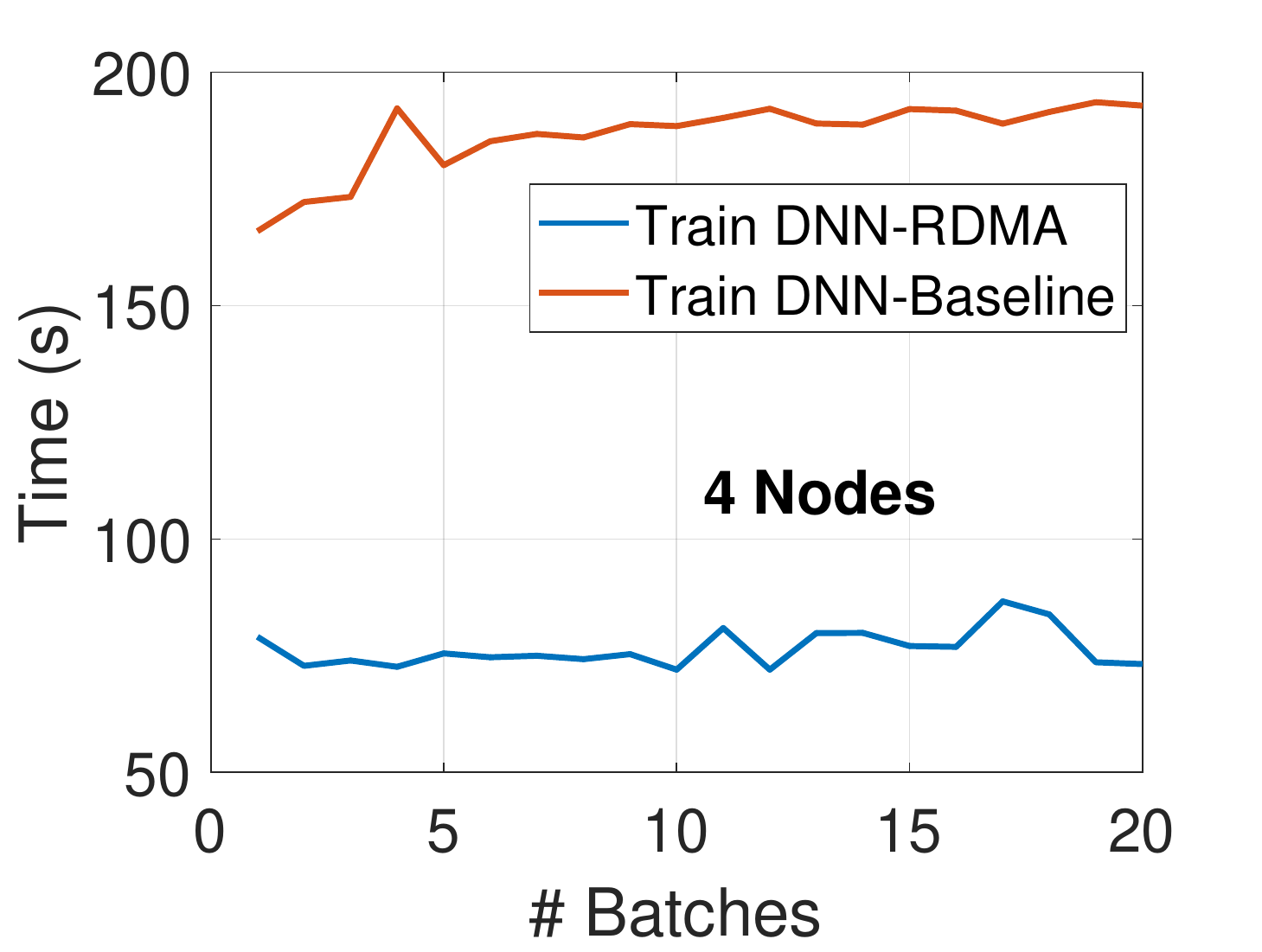}
}

\vspace{-0.15in}

\caption{Time of overall DNN training with/without RDMA.}
\label{fig:time-dnn-rdma}
\vspace{-0.3in}
\end{figure}

\noindent \textbf{GPUDirect RDMA.}
Figure~\ref{fig:rdma} shows the push communications time for 2 and 4 nodes.
We fix the total training data the same. With more workers introduced in more nodes, the total \#Batches is reduced. Therefore, the figure shows $40$ batches for 2 nodes and $20$ batches for 4 nodes.
As expected, the communications increase when more nodes are involved. The GPUDirect RDMA enables GPUs to communicate with GPUs on other nodes without involving CPUs. It reduces as much as $87\%$ communications for 2-node setting and $84\%$ for the 4-node case. As a result of the communication time reduction, the overall training time is reduced $54\%$ (Figure~\ref{fig:time-dnn-rdma}).

\begin{figure}[htbp]
\centering
\mbox{
	\includegraphics[width=2.2in]{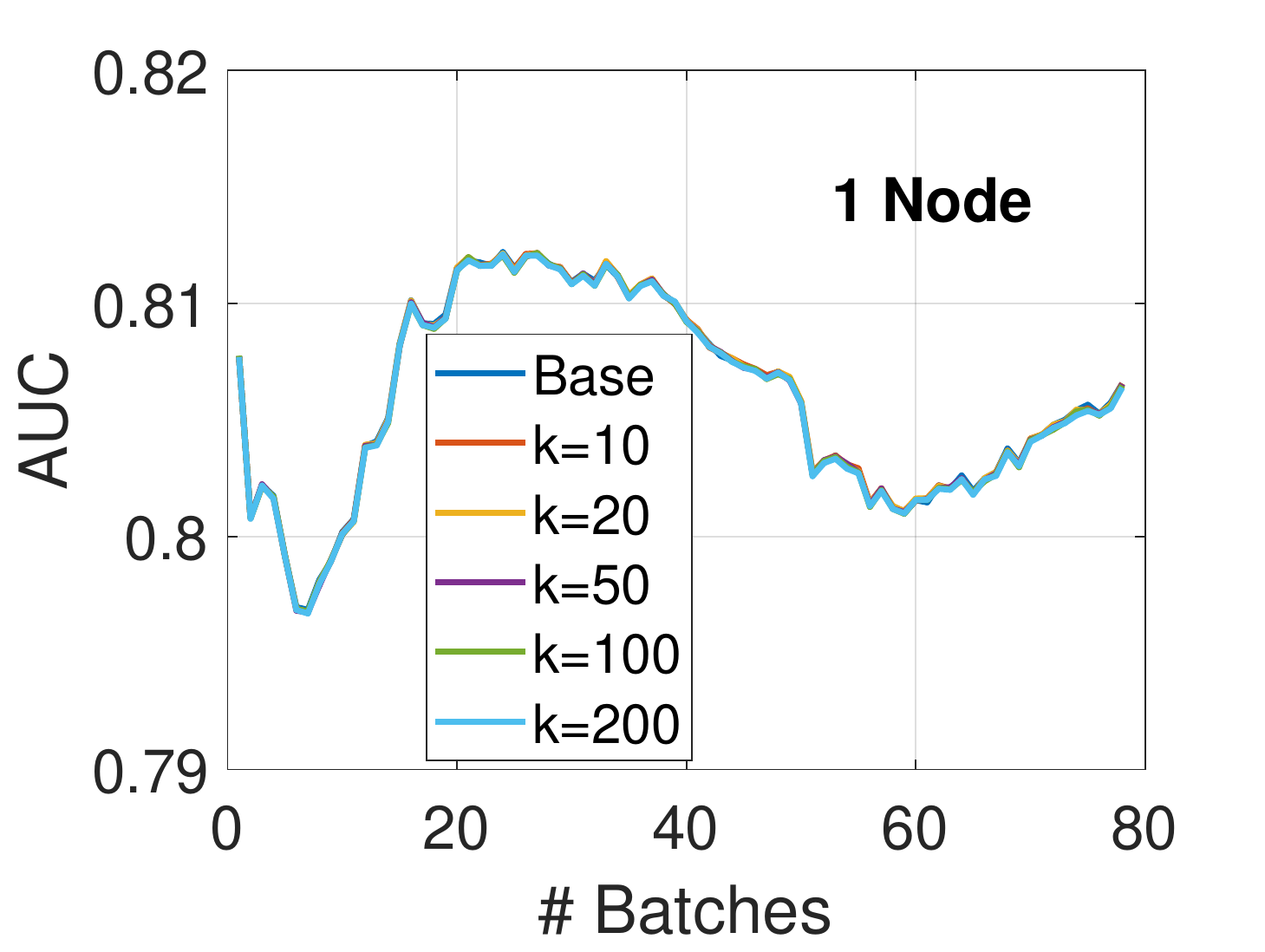}
	\hspace{0.15in}
	\includegraphics[width=2.2in]{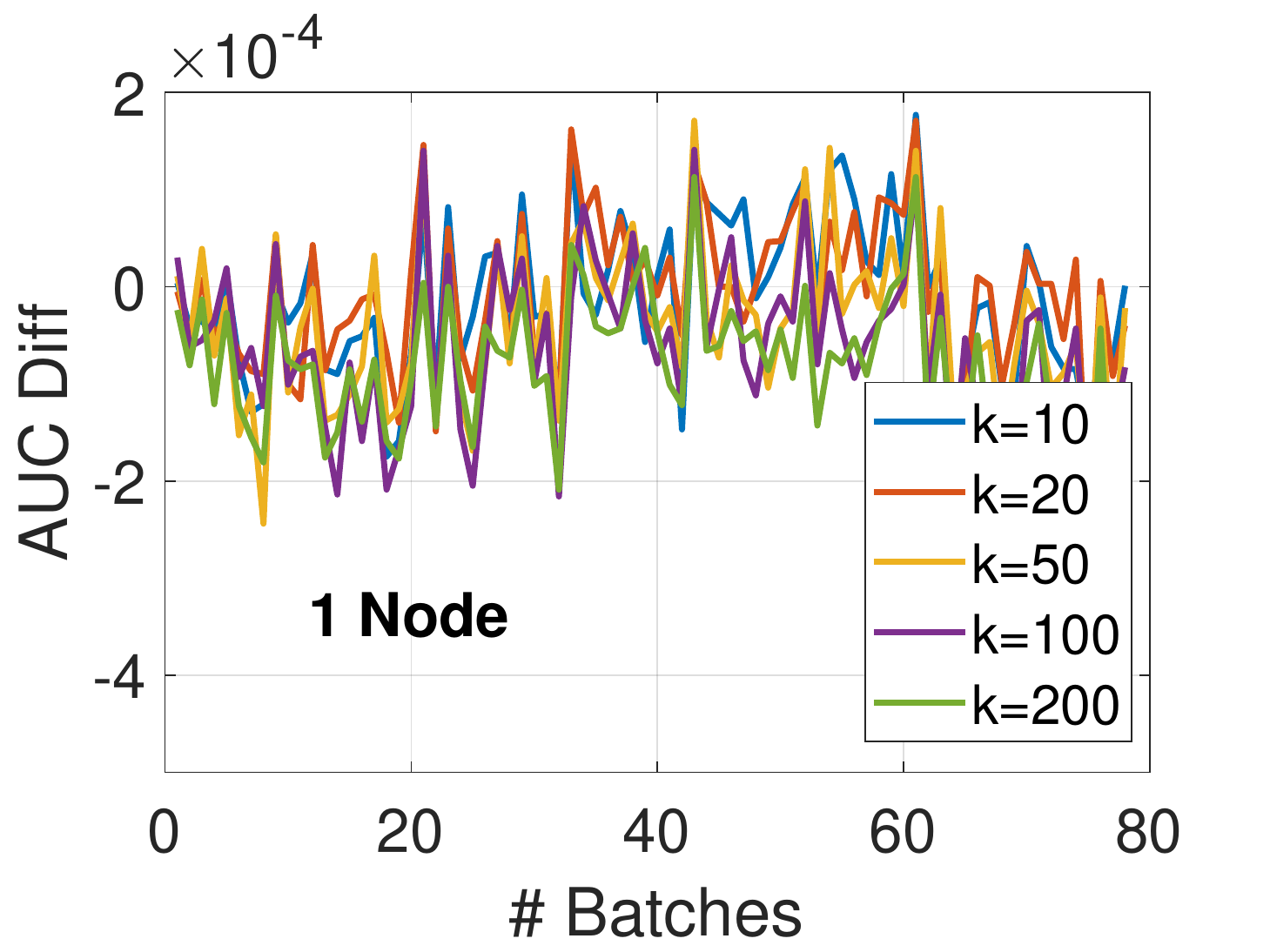}
}

\mbox{
	\includegraphics[width=2.2in]{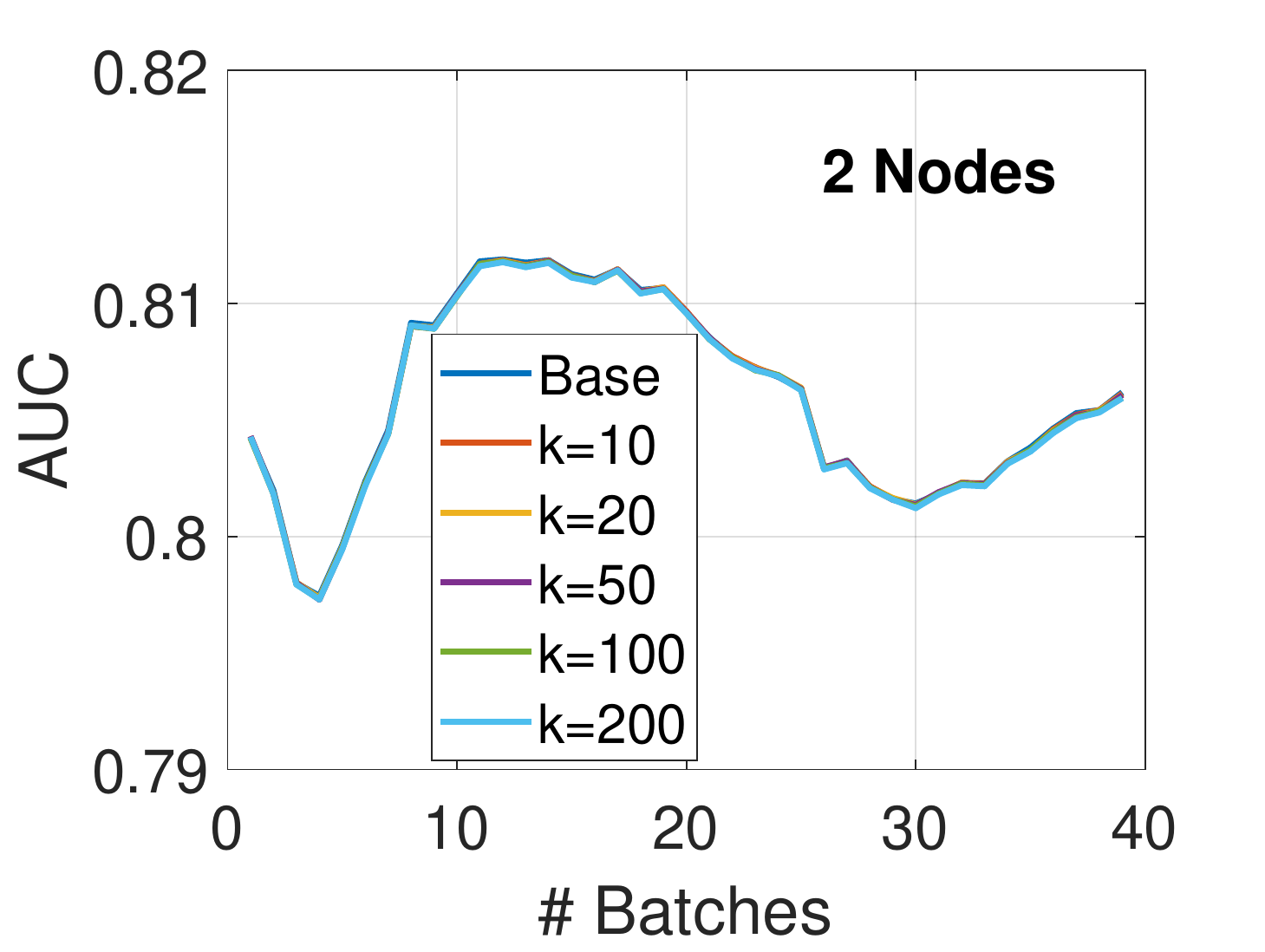}
	\hspace{0.15in}
	\includegraphics[width=2.2in]{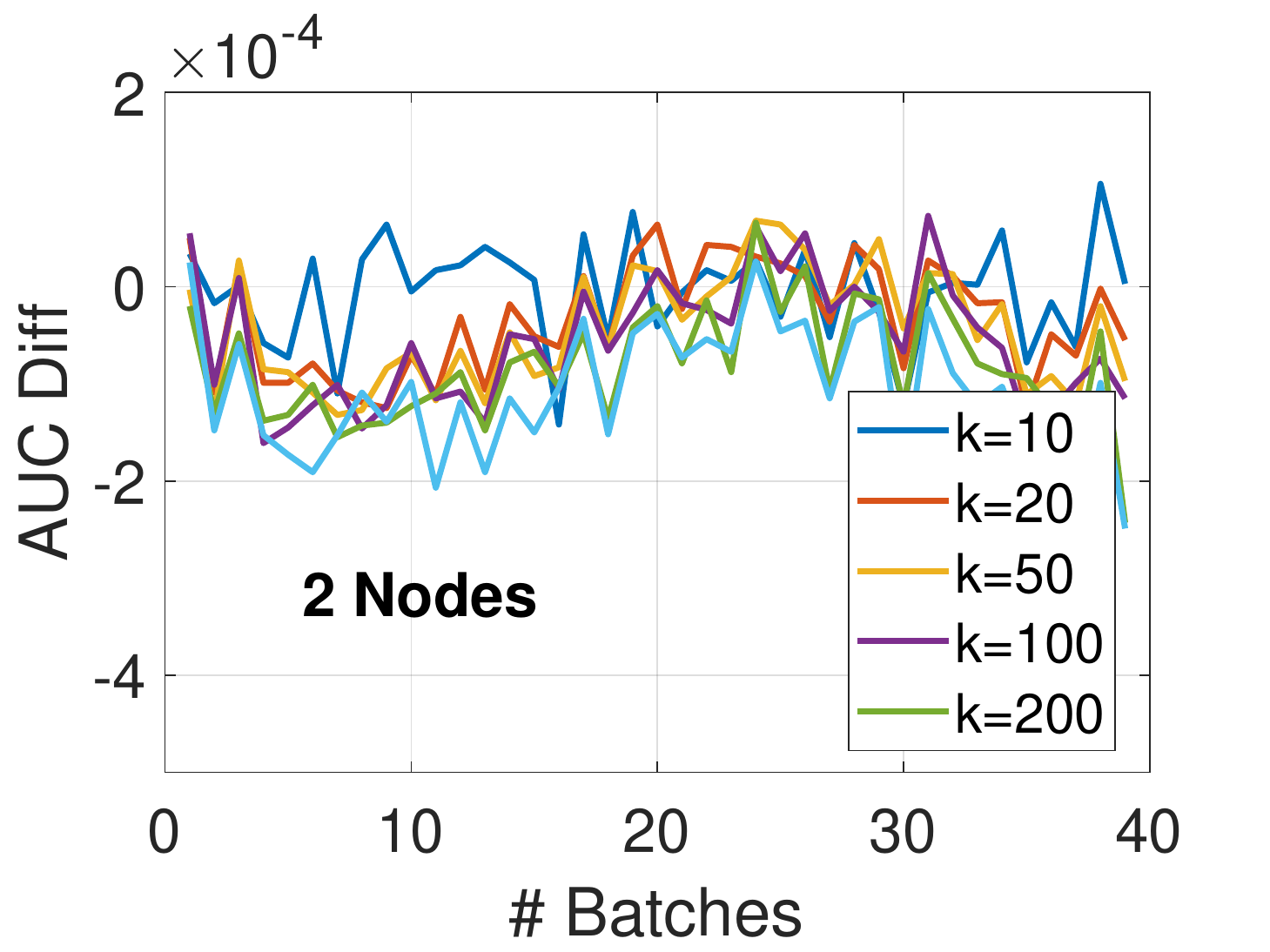}
}

\mbox{
	\includegraphics[width=2.2in]{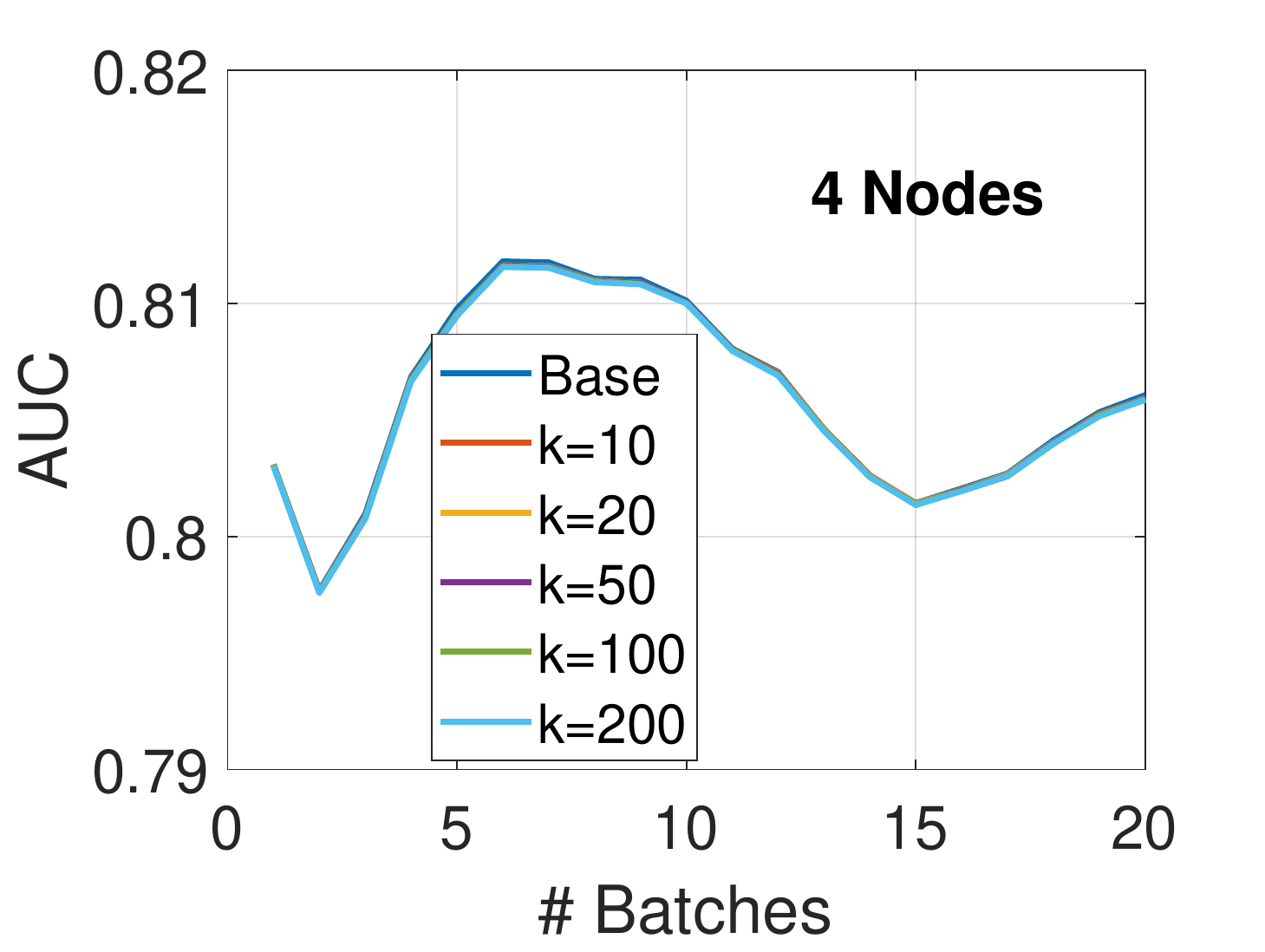}
	\hspace{0.15in}
	\includegraphics[width=2.2in]{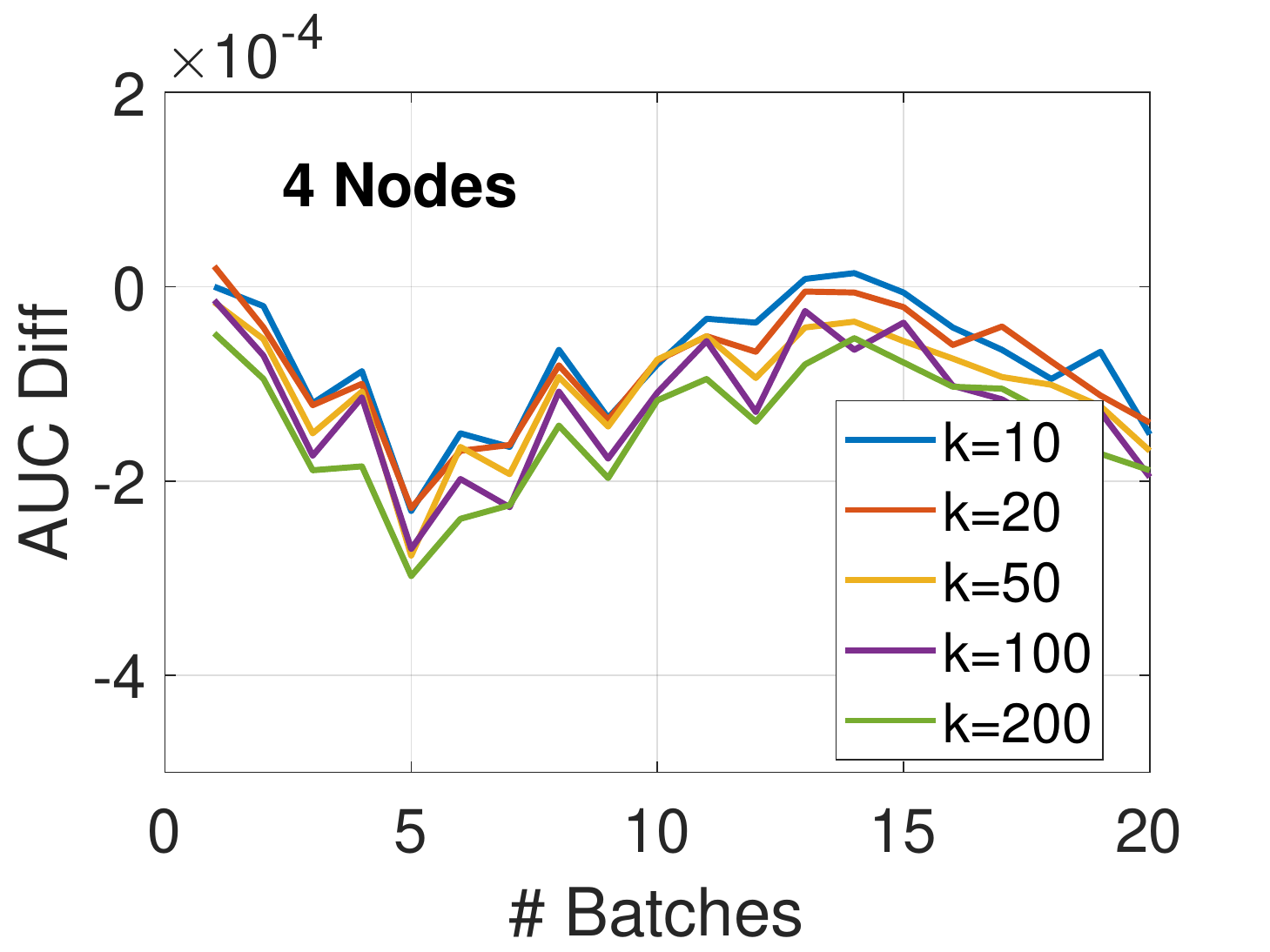}
}

\mbox{
	\includegraphics[width=2.2in]{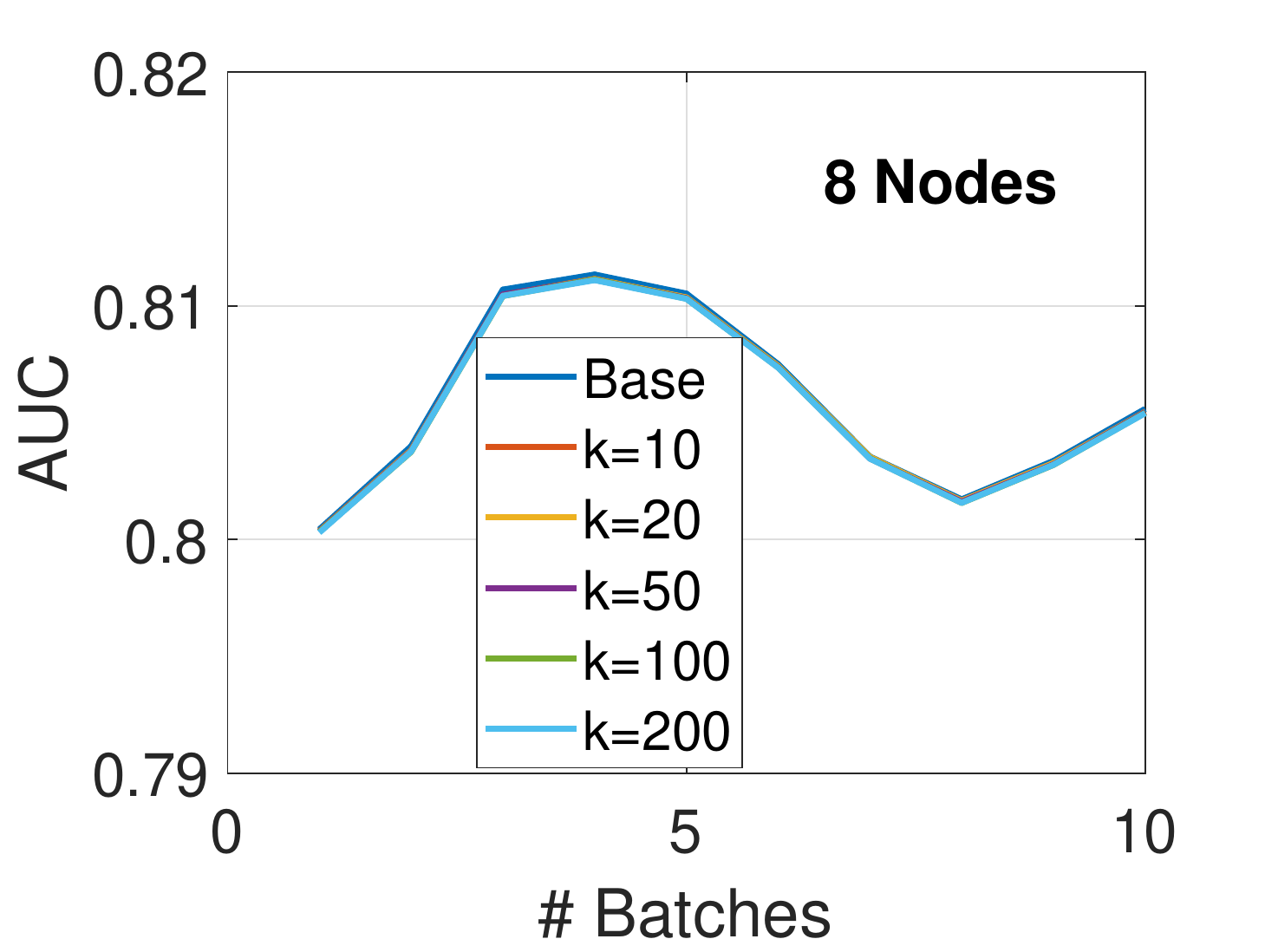}
	\hspace{0.15in}
	\includegraphics[width=2.2in]{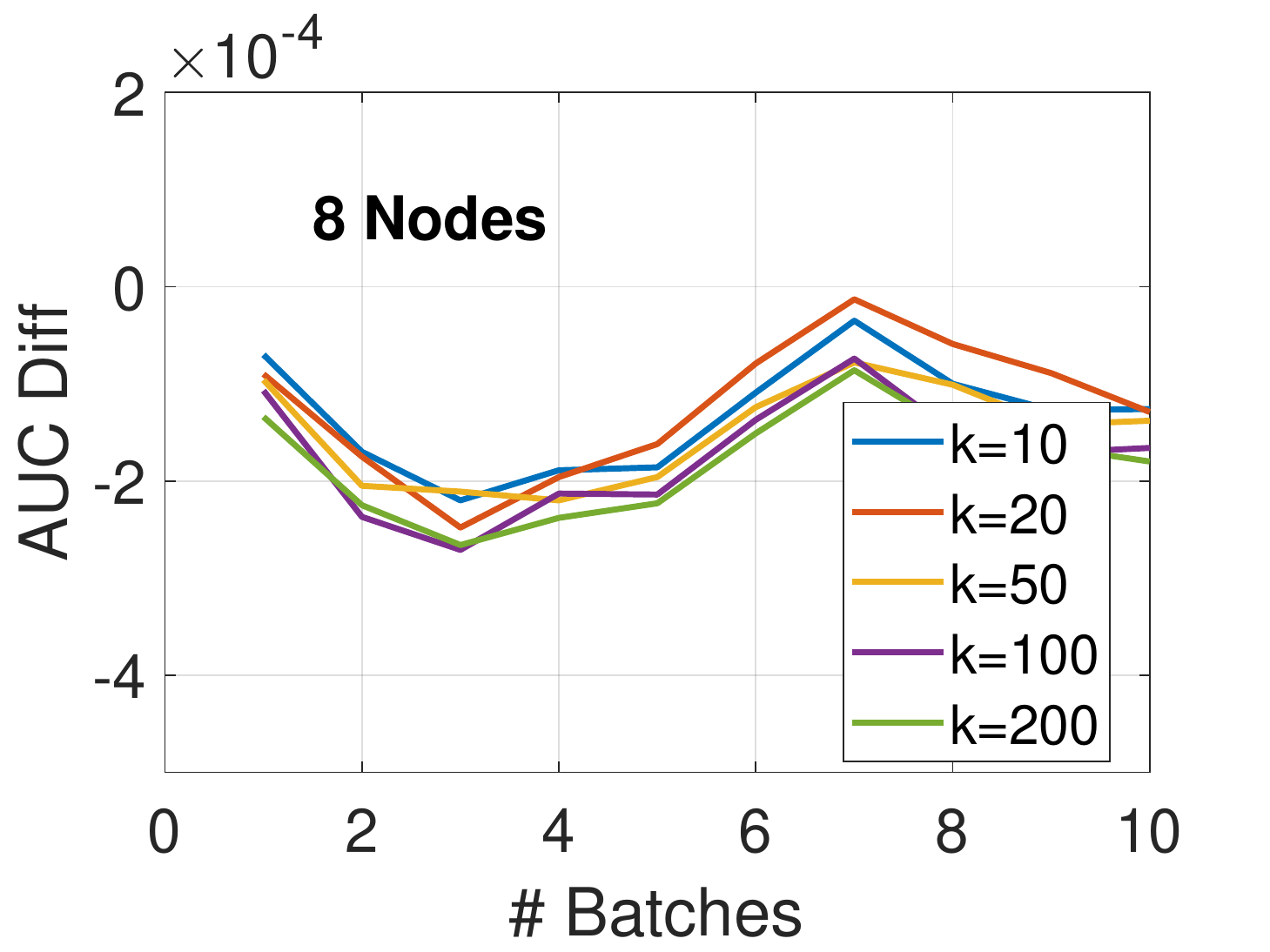}
}

\caption{$k$-step communication-efficient training. Left: AUC vs. number of training batches. Right: AUC difference of ($k$-step - baseline). }
\label{fig:auc}
%\vspace{-0.2in}
\end{figure}

\subsection{Communication of $k$-Step Model Merging}
The optimizations we investigated above only modify the data communication routes, which are lossless. However, the proposed $k$-step model merging touches the training algorithm. In this section, we evaluate our proposed $k$-step model merging algorithm in two measures: (a) accuracy; and (b) execution time.

\vspace{0.1in}
\noindent\textbf{Accuracy.}
The CTR prediction accuracy is one of the most critical measures for industry advertising applications. We have to guarantee no noticeable accuracy drop when we apply any optimizations. Here we employ Area Under the Curve~\citep{huang2005using} (AUC) as the quality measure of our trained models. Figure~\ref{fig:auc} depict the AUC for 1, 2, 4, and 8 node settings. The right part of the figure is a zoomed-in view of the left part. When we vary the $k$ from $10$ to $200$, almost no differences in AUC can be observed in the left part of the figure. In the online learning context, it is normal to observe a higher AUC in working hours since there are more training instances in those hours---the model can better predict the user behaviors in that time period. In the zoomed-in view, we can observe that the AUC diff vibrates within 0.0002\% range in most settings. We can assume the AUC loss is minimal and statistically ignorable.

\begin{figure}[htbp]
\centering
\mbox{
	\includegraphics[width=2.2in]{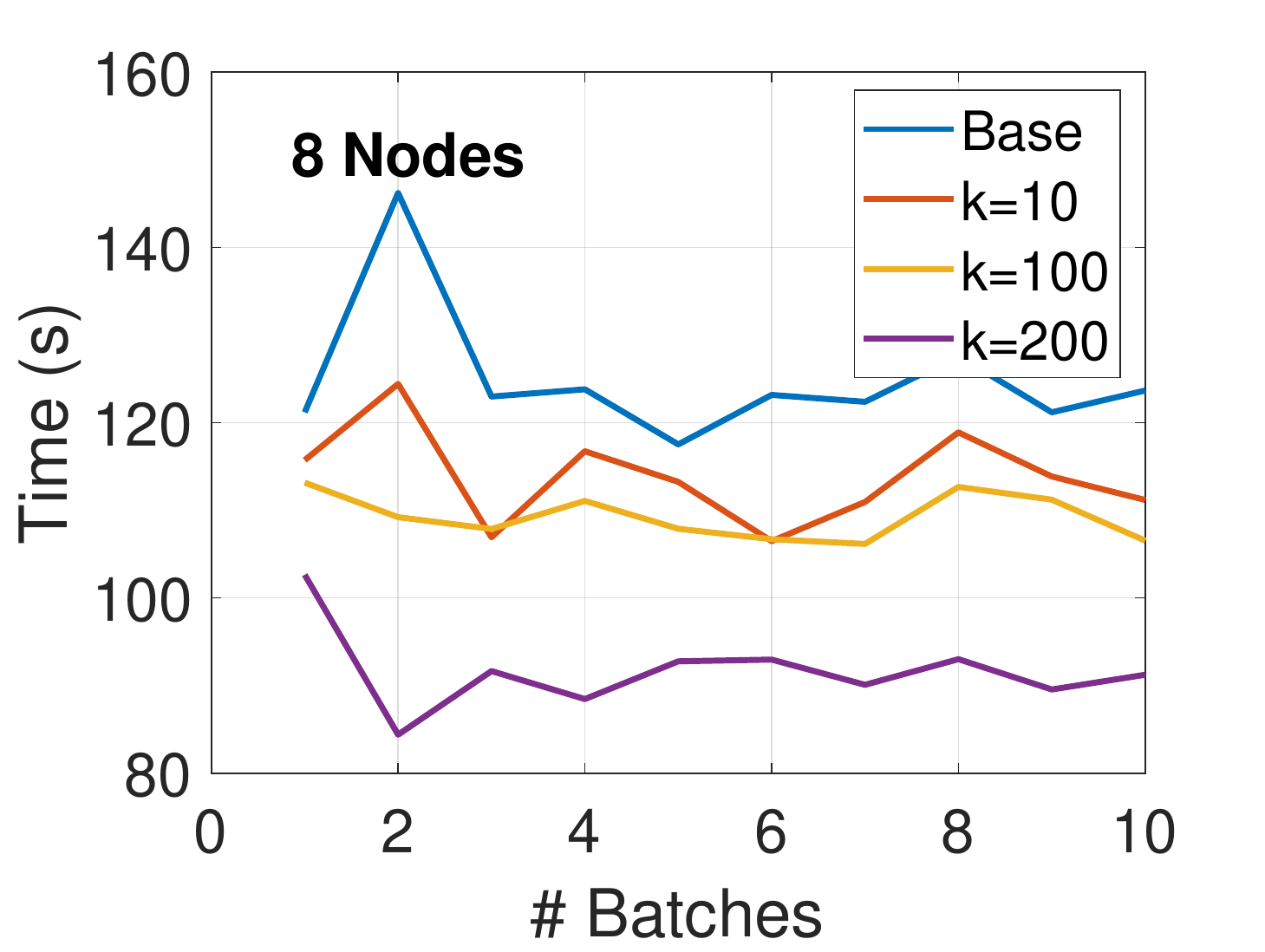}\hspace{0.15in}
	\includegraphics[width=2.2in]{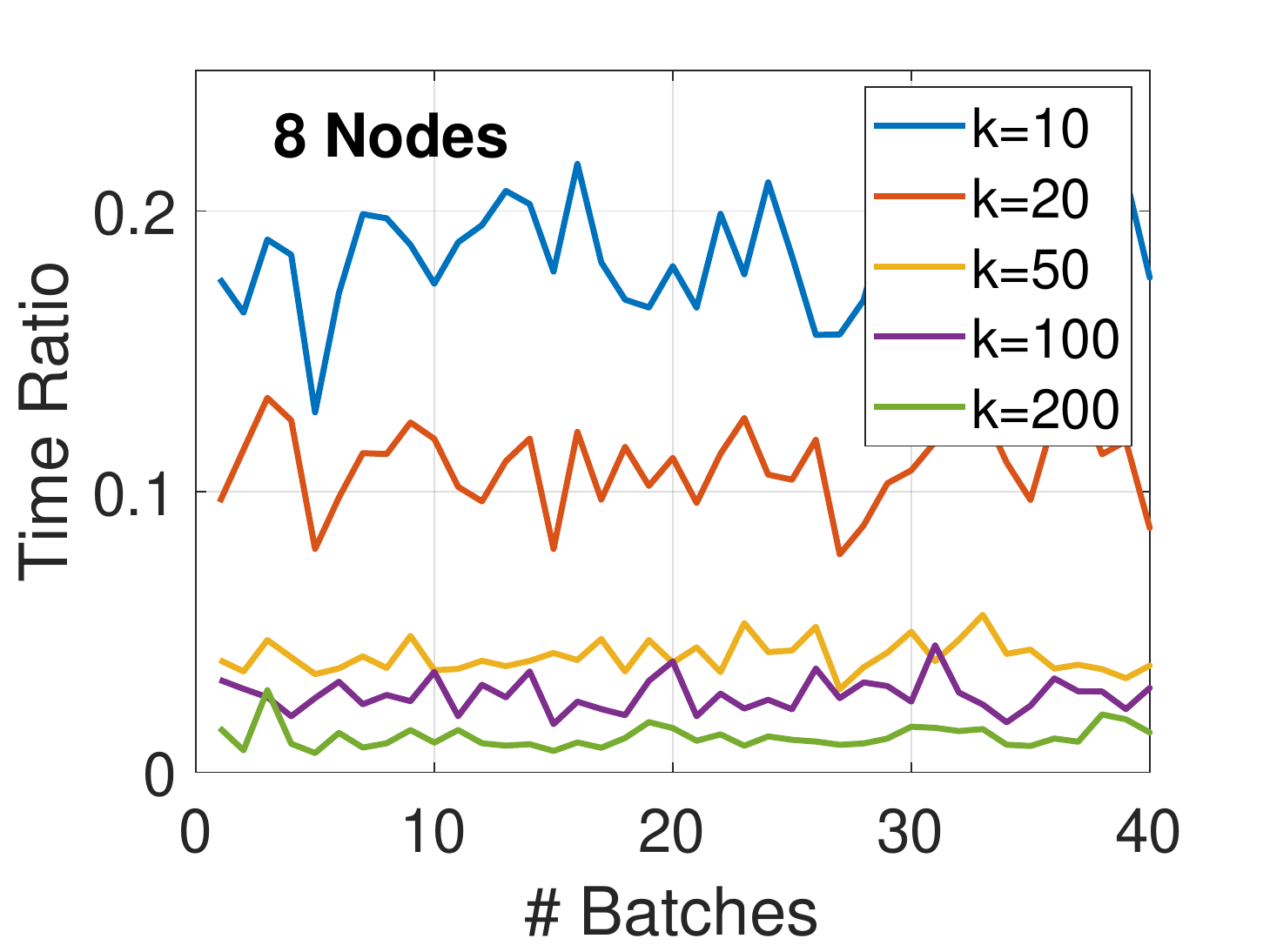}
}
\caption{Left: time of DNN training with $k$-step communication scheme. Right: Communication time (model transmission) ratio of $k$-step over baseline. The average ratio is $18.1\%,10.8\%,6.4\%,2.8\%$ and $1.2\%$, for $k=10,20,50,100,200$.}
\label{fig:k-step-time}
\end{figure}

\noindent\textbf{Execution time.}
Figure~\ref{fig:k-step-time} reports the overall execution time to train the model with/without $k$-step model merging on 8 nodes. We can reduce as much as 25\% execution time when $k=200$. The right part of the figure presents the communication time ratio of $k$-step over the baseline case (all optimizations except $k$-step). The communication time is approximately linearly reduced as we increase $k$. The overall execution time is not proportionally reduced because the training system is executed in a pipeline---there is synchronization overhead with other stages.

\subsection{Discussion}
Now we can answer the questions driven the experimental evaluation. Our proposed core binding and SSD I/O improve the sparse parameter pulling  stage around $15\%$. For inner-node GPU communication, our two-phase GPU communication algorithm reduces around $90\%$ GPU communication time. For the inter-node case, our GPUDirect RDMA optimization cuts the communication cost to 1/6 and 1/8 for 2-node and 4-node settings, respectively. In terms of accuracy, through 1-node to 8-node different scenarios, our proposed $k$-step merging algorithm is able to generate almost the same accurate result comparing with the original optimizer---the AUC diff is within 0.0002\% range. Meanwhile, our proposed training framework is more than 4X faster to finish the training execution.

\section{Conclusion}
In this paper, we propose a hardware-aware training workflow that couples the hardware topology into the algorithm design to improve the communication bandwidth between devices. To reduce the extensive communication between computing nodes, we introduce a $k$-step model merging algorithm for Adam and theoretically provide its convergence rate in non-convex optimization. We evaluate the proposed system on real-world ads data. The empirical results show that our proposed optimizations improve both the inner-node communication bandwidth and the inter-node communication.
The experimental evaluation confirms that our proposed training framework achieves better execution time while maintaining the same test accuracy.

%\begin{acks}
%\end{acks}

%\newpage
%\clearpage

\bibliography{mlsys}
\bibliographystyle{plainnat}

\end{document}